\documentclass[graybox]{svmult}

\usepackage{type1cm}        
\usepackage{makeidx}         
\usepackage{graphicx}        
\usepackage{multicol}        
\usepackage[bottom]{footmisc}

\usepackage{newtxtext}       %
\usepackage{newtxmath}       
\usepackage{listings}
\usepackage{pifont}
\usepackage{multirow}
\usepackage{tablefootnote}
\usepackage{hyperref}
\usepackage[frozencache,cachedir=.]{minted}
\AtBeginEnvironment{minted}{%
  }

\usepackage[
    group-digits=integer, group-minimum-digits=4, 
    free-standing-units, unit-optional-argument, 
    ]{siunitx}
\usepackage{pgfplots}
\pgfplotsset{
    compat=1.9,
    log ticks with fixed point, 
    table/col sep=tab, 
    unbounded coords=jump, 
    filter discard warning=false, 
    }

\makeindex             

\begin{document}

\title*{Klever: Verification Framework for Critical Industrial C Programs}

\author{Ilja Zakharov, Evgeny Novikov, Ilya Shchepetkov}

\institute{This work was carried out while authors were working at Ivannikov Institute for System Programming of the Russian Academy of Sciences, Moscow, Russia\\
\email{{ilja.s.zakharov, eunovm, ilya.shchepetkov}@gmail.com}}

\maketitle

\abstract{
Automatic software verification tools help to find hard-to-detect faults in programs checked against specified requirements non-interactively.
Besides, they can prove program correctness formally under certain assumptions.
These capabilities are vital for verification of critical industrial programs like operating system kernels and embedded software.
However, such programs can contain hundreds or thousands of KLOC that prevent obtaining valuable verification results in any reasonable time when checking non-trivial requirements.
Also, existing tools do not provide widely adopted means for environment modeling, specification of requirements, verification of many versions and configurations of target programs, and expert assessment of verification results.
In this paper, we present the Klever software verification framework, designed to reduce the effort of applying automatic software verification tools to large and critical industrial C programs.
}


\section{Introduction}
Tools that implement an automatic software verification technique also known as software model checking aim at finding violations of specified requirements non-interactively and proving program correctness under certain assumptions formally~\cite{Jhala:2009:SMC}.
Automatic software verification tools (for brevity, we will refer to them as \textit{verification tools} below) participating in annual competitions on software verification (\mbox{SV-COMP}) demonstrate excellent results for verification tasks included in the benchmark suite~\cite{Beyer:2021:SVC}.
Some verification tools miss a few faults and report fewer false alarms.
The benchmark suite contains moderate-sized programs that are prepared in advance and checked against the predefined list of properties.
However, these verification tools cannot be applied to large software projects without prior preparation.

This paper considers the verification of large critical industrial C programs such as operating system kernels, embedded software, web servers, database management systems, libraries and utilities on which many people rely in their daily lives and at work.
Such programs can contain hundreds or thousands of KLOC and might evolve rapidly (Table~\ref{tab:programExamples}).
Neither advanced algorithms implemented by developers of verification tools nor a substantial increase in computing power can guarantee obtaining valuable verification results in a reasonable time if large enough parts of these programs' source code are strongly relevant to checked properties.

\definecolor{gitgreen}{rgb}{0.0, 0.5, 0.0}

\begin{table}
\caption{Characteristics of several open-source programs\tablefootnote{Most of the source code of these programs is written in C. Data was collected using \textit{cloc} and \textit{git}.}.}
\label{tab:programExamples}
\centering
\begin{tabular}{ | l  | c | c | }
\hline
\textbf{Program name and version} &
\textbf{Size (MLOC)} &
\textbf{Lines changed\tablefootnote{Number of lines added and removed during a period of one year before the release. These numbers were divided by total sizes of programs to exemplify a pace of changes.} (MLOC)} \\
\hline
Linux 5.7.7 kernel & 27 & \begin{tabular}{@{}c@{}} \textcolor{gitgreen}{2.6 added (\SI{9.7}{\percent})} \\ \textcolor{red}{1.2 deleted (\SI{4.4}{\percent})} \end{tabular} \\
\hline
FreeBSD 12.1 & 25 & \begin{tabular}{@{}c@{}} \textcolor{gitgreen}{5.3 added (\SI{22}{\percent})} \\ \textcolor{red}{3 deleted (\SI{12}{\percent})} \end{tabular} \\
\hline
GCC 10.1 & 14 & \begin{tabular}{@{}c@{}} \textcolor{gitgreen}{2.2 added (\SI{15}{\percent})} \\ \textcolor{red}{0.99 deleted (\SI{6.9}{\percent})} \end{tabular} \\
\hline
GTK 3.24.21 & 3.2 & \begin{tabular}{@{}c@{}} \textcolor{gitgreen}{0.13 added (\SI{4.2}{\percent})} \\ \textcolor{red}{0.1 deleted (\SI{3.2}{\percent})} \end{tabular} \\
\hline
PostgreSQL 12.3 & 2.2 & \begin{tabular}{@{}c@{}} \textcolor{gitgreen}{0.13 added (\SI{5.3}{\percent})} \\ \textcolor{red}{0.07 deleted (\SI{3.3}{\percent})} \end{tabular} \\
\hline
glibc 2.31 & 2 & \begin{tabular}{@{}c@{}} \textcolor{gitgreen}{0.24 added (\SI{12}{\percent})} \\ \textcolor{red}{0.25 deleted (\SI{13}{\percent})} \end{tabular} \\
\hline
\end{tabular}
\end{table}

The interest in the application of verification tools to critical industrial programs continually grows due to the following reasons:
\begin{itemize}
\item Faults in these programs can result in significant economic losses~\cite{Brooks:1987:NSB,Tassey:2002:EII}.
\item Traditional software quality assurance techniques such as code review, testing and static analysis often cannot provide correctness guarantees~\cite{Black:2016:DRS,Glass:2002:FFS}.
\item Heavyweight formal verification methods like deductive verification are capable of proving functional correctness under certain assumptions.
But such methods are successfully applied either to fairly small programs whose size is up to 12~KLOC~\cite{Alkassar:2010:PVO,Gu:2015:DSC,Klein:2014:CFV} or to critical components of large programs~\cite{Efremov:2017:DVU,Ferreira:2014:AVF,Gotsman:2011:MVP} since verification of them as a whole would require enormous efforts.
\end{itemize}

Target programs can extensively interact with their environment during operation, while verification tools are unaware of any restrictions imposed on the corresponding interactions.
Users may need to check that their programs satisfy specific requirements in addition to supported properties.
Program environment and requirements can vary for different versions and configurations of target programs.
However, there is a lack of widely adopted methods and tools for preparing programs for large-scale verification.
At last, existing verification tools do not provide users with a comprehensive enough suite of means for expert assessment of verification results.

To reduce efforts that are necessary for the application of automatic software verification tools for large industrial C programs, we have been developing the Klever software verification framework and project-specific adaptations considered in Sections~\ref{section:klever} and \ref{section:adaptations} correspondingly.
Section~\ref{section:results} demonstrates the publicly available achievements of Klever and estimates various aspects of its usage.
Section~\ref{section:relatedwork} outlines the related work.
In conclusion, we summarize our overall vision of the practical application of automatic software verification tools.

The main contribution of the paper is an approach and its implementation in the Klever verification framework, tailored to apply verification tools to large industrial programs.
This approach is a complex set of methods that tackle each step of a path, from selecting the scope of the source code to be checked to analyzing verification results.
We provided a solution for each significant challenge that a user might face on this way.
Consequently, the paper might look too broad and cover too many aspects.
But focusing on a single individual problem, like environment modeling or program decomposition, does not allow seeing the big picture and simplifying the overall verification process in practice.
Therefore, our main contribution is a solid approach to the application of automatic software verification tools to industrial C programs that breaks down into corresponding methods for:
program decomposition;
environment modeling;
stating requirement specifications;
presenting verification results and enabling their triage.

In addition, the presented verification framework can work with any verification tool that supports its interface, which is pretty close to the one adopted by the \mbox{SV-COMP} community.
The most vital requirements for us are the possibility to apply the tool directly to the source code, get verification results non-interactively, and visualize them.
We stick to these criteria mostly for technical reasons and do not limit the variety of verification approaches and techniques that such tools can implement under the hood.
Discussing the application of other analysis and verification tools to various software is not our goal in the paper as it is too broad, so we limit the overview of related work to the particular problem of applying the aforementioned tools to industrial programs.

\section{Klever: A Software Verification Framework}
\label{section:klever}

\begin{figure}
    \centering
    \includegraphics[width=1\textwidth]{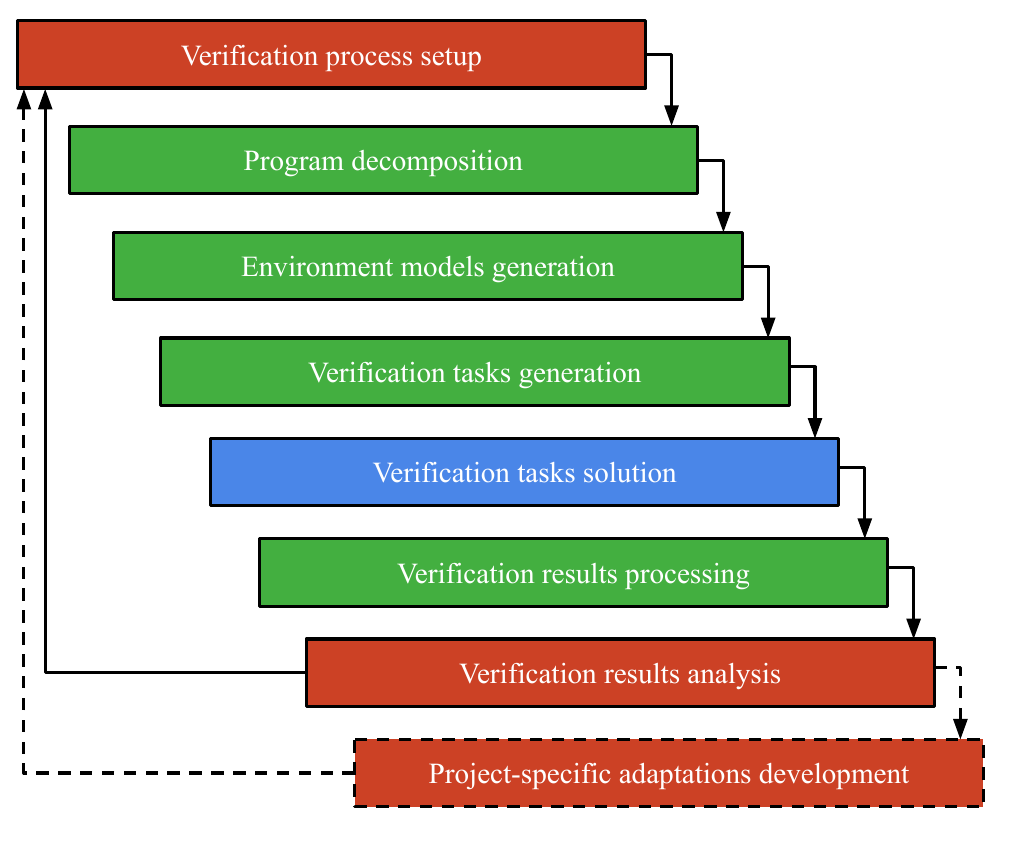}
    \caption{Klever workflow.}
    \label{fig:klever}
\end{figure}

Fig.~\ref{fig:klever} presents the overall workflow of Klever~\cite{Novikov:2018:TAS}.
Rectangles filled with red represent manual actions.
Green rectangles mean automatic steps executed by Klever.
The only blue rectangle corresponds to the invocation of verification tools.
The rectangle with a dotted border represents an optional step necessary if users would like to improve verification results.

The Klever software verification framework uses \mbox{SV-COMP} compliant verification tools as backends and automates the following steps (they are considered in detail in corresponding subsections below):
\begin{itemize}
\item Decomposition of programs into fragments of a moderate size.
\item Generation of environment models for various kinds of interactions with an environment.
\item Generation of verification tasks that includes configuration of verification tools.
\item Scheduling and monitoring to enable parallel generation and solution of verification tasks.
\item Preliminary processing of verification results.
\item Managing verification processes and expert assessment of verification results.
\end{itemize}

Klever supports verification of programs developed in the GNU~C programming language on \textit{x86\_64}, \textit{ARM}, and \textit{ARM64} Linux platforms.

Project-specific adaptations include configurations, specifications and plugins for checking specific software.
If a project adaptation has already been implemented, users need to deploy Klever and can start using it as is without any additional manual efforts.
Otherwise, several iterations dedicated to the development of the project-specific adaptation are necessary.

During design and development, we cared about supporting the verification of different versions and configurations of target programs.
This is highly demanded by the fast-changing industry.
Klever does not introduce any changes to the program's source code or build processes.
Thus, it can pick up a new version or configuration of a program smoothly to run its verification without extra steps except for adjusting an appropriate project-specific adaptation if necessary.

The Klever software verification framework is an open-source project\footnote{https://forge.ispras.ru/projects/klever, https://github.com/ldv-klever/klever} primarily implemented in Python~3.
Configurations from project-specific adaptations are stored as JSON files.
Specifications are developed using appropriate domain-specific languages.
Project-specific adaptation plugins are implemented as Python~3 modules.
The Klever project repository contains numerous tests written in C.
These tests are checked automatically by a CI/CD system before any substantial update to the main branch.
Klever user documentation can be found at this link\footnote{https://klever.readthedocs.io/en/latest}.

The following subsections describe particular steps of the automated Klever workflow and related components of the Klever software verification framework.
It is worth noting that different Klever components depend on several auxiliary tools considered in Appendix~\ref{Appendix:AuxiliaryTools}.
One can omit any of these subsections if they are not of interest.

\subsection{Decomposition of Programs}

One can hardly verify any complicated, large industrial program automatically as one piece of code.
We suggest decomposing large programs by extracting smaller \textit{program fragments} from their source to verify them separately.
Such an approach decreases the demand for computational resources and simultaneously increases the efforts required for environment modeling  (more on this in the next subsection).

We suggest considering logically interconnected program components like, say, loadable kernel modules or plugins (Table~\ref{tab:components}) as program fragments.
The decomposition is performed only on the file level.
Each program fragment can be considered as a set of C source files.
In our experience, this is a "golden mean" that enables obtaining useful verification results with moderate efforts for modeling the environment.
Each program fragment should have at least one entry point --- a function that can be called by other program fragments.

\begin{table}
\caption{Approximate number of components in open-source projects.}
\label{tab:components}
\centering
\begin{tabular}{ | l | c | }
\hline
\textbf{Project} & \textbf{Number of components (approximately)} \\
\hline
Linux Kernel & \SI{5000}{} \\
\hline
BusyBox & 300 \\
\hline
GTK & 200 \\
\hline
Apache & 150 \\
\hline
VLC & 80 \\
\hline
\end{tabular}
\end{table}



\begin{figure}
  \centering
    \includegraphics[width=1\textwidth]{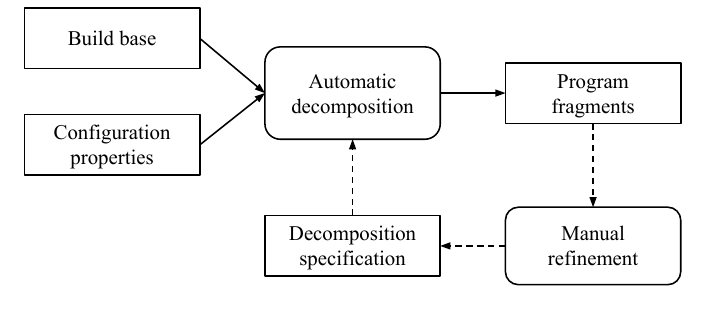}
    \caption{Iterative method of program decomposition.}
    \label{figure:decomposition}
\end{figure}

It is hardly possible to propose a universal algorithm for an automatic program decomposition.
Thus, we suggest a configurable and extendable approach illustrated in Fig.~\ref{figure:decomposition}.
A user should provide configuration properties and the program's source code in the form of a Clade build base\footnote{One can see Appendix~\ref{Appendix:Clade} for more details about Clade and build bases.}.
Program decomposition is performed automatically.
However, a user can adjust it by tuning configuration properties and the decomposition specification between runs of Klever.

The Project Fragment Generator (PFG) is a Klever component that performs the automatic decomposition.
It implements the algorithm illustrated in Fig.~\ref{listing:pfg_pseudocode}.
Configuration properties, which are denoted as \textit{conf}, contain the names of files, directories and program fragments for verification.
It also lists the names of tactics that implement particular decomposition~(\textit{decomposition\_tactic}) and composition~(\textit{composition\_tactic}) algorithms.
These tactics ought to be implemented as part of a specific program adaptation, and their implementation should be based on a particular program architecture.
However, the repository contains several simple open-source implementations of such tactics.

The build base~(\textit{build\_base}) contains, among other things, an oriented callgraph and build command graph, which are used by PFG\footnote{The callgraph consists of nodes that correspond to functions and edges that match function calls. An edge of a build command graph corresponds to a file that was produced by some build command and then provided to another build command as an input. Nodes in this graph reflect build commands.}.
The build base contains other details to resolve dependencies between files, such as definitions of macros and types.
We focus mostly on the callgraph because functions are essential means for implementing the main logic of the program, and the distribution of functions among files better reflects how the program is organized.

\begin{figure}
\begin{minted}[frame=single,fontsize=\scriptsize]{python}
def PFG(conf, build_base):
    decomposition_tactic, composition_tactic = get_methods(conf)
    decomposition_spec = get_decomposition_spec(conf)
    final_program_fragments = set()

    file_graph = prepare_file_graph(build_base)
    fragments_graph = decomposition_tactic(conf, build_base, file_graph)
    if decomposition_spec:
        fragments_graph = refine_fragments(fragments_graph, decomposition_spec)
    target_files = resolve_files(fragments_graph, conf, build_base)
    target_fragments = resolve_fragments(fragments_graph, target_files)

    for fragment in target_fragments:
        extra_fragments = composition_tactic(fragment, fragments_graph, conf,
                                             build_base, file_graph)
        new_target = Fragment(fragment.files)
        for extra_fragment in extra_fragments:
            new_target.add_files(extra_fragment.files)
            final_program_fragments.add(new_target)

    return final_program_fragments
\end{minted}
\caption{Algorithm of decomposition into program fragments.}
\label{listing:pfg_pseudocode}
\end{figure}

First, PFG reads \textit{conf} and chooses two project-specific tactics (for decomposition and composition) depending on a program, its version and names provided by a user.
The user should choose the best suitable tactics for decomposition and composition, depending on the program's design and its build process.
Decomposition and composition tactics traverse the build command graph and callgraph selecting files for program fragments for decomposition or merging already selected program fragments for composition.

The decomposition specification (\textit{decomposition\_spec}) is a JSON file that specifies files to be added or removed from particular program fragments\footnote{There are a couple of examples of decomposition specifications in Appendix~\ref{Appendix:Fragments}.}.
PFG uses it in the next step.

Then PFG constructs a \textit{file graph} using the two mentioned graphs.
Nodes of the file graph correspond to the program's C source files, and edges match function calls between them.
Then, the \textit{decomposition\_tactic} constructs a \textit{program fragment graph} (\textit{fragments\_graph}) for the particular program under verification.
Each node of this graph corresponds to a set of C source files of the program.
An edge between two program fragments exists if the first one calls at least one function defined in the second one.

Decomposition and composition tactics work with the file graph, callgraph and build command graph.
Common operations with them, such as traversing over dependencies or finding specific nodes, are separated into a specific shared library that simplifies the implementation of new tactics.
One can find more information about the configuration of program decomposition in the user documentation\footnote{https://klever.readthedocs.io/en/latest/dev\_decomposition\_conf.html}.

The \textit{refine\_fragments} step is the correction of the generated program fragments according to the provided decomposition specification.
This specification serves only the purpose of manually adjusting the automatically obtained results of program decomposition using tactics.
To achieve that, the user can modify the generated decomposition specification by changing the list of functions, files, and directories (either concrete names or regular expressions) describing program fragments.
This specification can be provided at the follow-up run of Klever, and it supersedes program fragments generated automatically.

At the next step, PFG marks nodes of the file graph and program fragment graph as target ones using the configuration.
A user can provide a list of functions, files, and directories (yet again as either concrete names or regular expressions) to specify targets for verification.
If a file is marked as a target one, then all program fragments that contain the file are also marked as targets.

The last step is the composition of program fragments.
The idea is to find a few other program fragments for each target program fragment and combine them together as new program fragments.
It might reduce efforts required for environment modeling.
For example, a composition tactic can search for program fragments that implement missing function definitions for some target program fragment.

\begin{figure}
  \centering
    \includegraphics[width=1\textwidth]{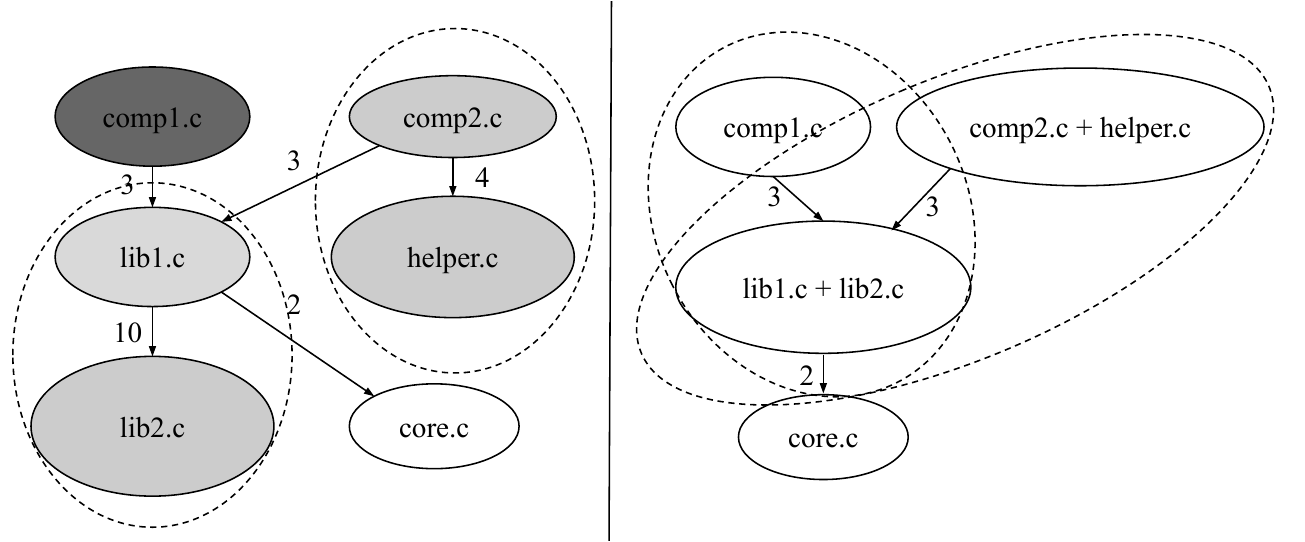}
    \caption{Decomposition and composition example.}
    \label{figure:decomposition_example}
\end{figure}

An example of decomposition and composition is shown in Fig.~\ref{figure:decomposition_example}.
There is a file graph in the left part of the picture.
For illustrating purposes, we added information about function calls (numbers), build command dependencies (shapes with the same color are input files of a single linking command) and lines of code per file (sizes of ellipse shapes).
There is a program fragment graph on the right side of the picture.
There are two target components (\textit{comp1.c} and \textit{comp2.c}) that actively use a couple of libraries (\textit{lib1.c} and \textit{lib2.c}).
The second component is implemented with a \textit{helper}; the first library refers \textit{core.c}.
Decomposition and composition tactics can use this information to simplify algorithms and make them more precise.
Imagine a user expects to get two program fragments for corresponding components with included dependencies for verification as a result of the decomposition.
PFG makes two program fragments, illustrated in the left picture, and then the composition tactic adds a program fragment with libraries to each component.
A dashed line on the right illustrates the result of such a composition.

It is worth mentioning that the process of decomposition is not very stable regarding changes in the code base if it relies on manual file or function listings.
However, if there is a way to automatically select program components and implement the routine as a decomposition tactic, then this approach becomes much more stable.
The Linux kernel is a good example.
It is hardly possible to manually list files that correspond to loadable kernel modules that might change significantly from version to version.
But an algorithm that detects files from the build commands tree allows seamlessly obtaining the list of files for each independent module, until the whole algorithm for arranging build commands is unchanged.

\subsection{Environment Modeling}


Libraries, user inputs, other programs, etc. constitute an environment that influences program execution.
Program verification requires providing a model that represents certain assumptions about the environment:
\begin{itemize}
\item It should contain a function that we refer to as an \textit{entry point}, and all paths analyzed by a verification tool should start with this function.
This entry point should invoke the target program's API in the same way as the real environment.
\item It should contain models of  functions that the program calls during execution and which can influence verification results, but not necessarily all of them.
These functions can be defined in other program fragments or libraries.
Moreover, some functions that require modeling could be just too complicated for verification tools and need to be replaced by simplified models.
\item It should correctly initialize external global variables.
\end{itemize}

Our experience shows that bug finding is possible even without accurate environment models if providing verification guarantees is out of scope.
However, it is crucial to provide precise environment models considering the specifics of checked requirements and programs under verification to achieve high-quality verification results~\cite{Zakharov:2018:CEM}.
It becomes even more important to do that to avoid missing faults and false alarms after program decomposition.

The Klever framework allows the development of the environment model either directly in the C programming language or using the proposed DSL approach.
A user might choose an approach that suits the project best.
We focus mostly on the DSL approach in this section, but before diving deeper into details, let us highlight the main advantages of the proposed solution.
The configurable DSL enables detaching the environment model from the code and maintaining a certain level of abstraction from program interfaces implemented by a particular program fragment. The approach pays off if there are many program fragments with a similar design, like operating system device drivers.
Otherwise, the development cost using the DSL might be close to the cost of developing the model in C manually, especially if each program fragment has an entirely unique interface and  logic.



\begin{figure}
  \centering
    \includegraphics[width=1\textwidth]{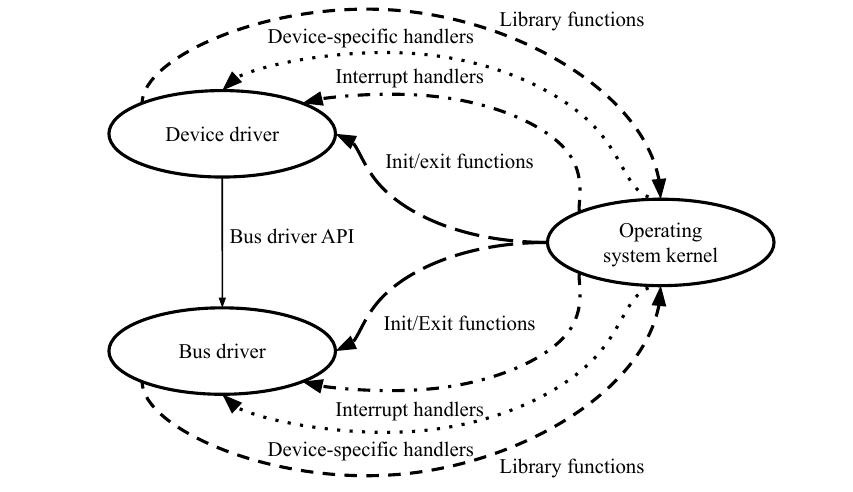}
    \caption{Interaction scenarios between drivers and their environment.}
    \label{figure:complex_interaction}
\end{figure}

We show different kinds of interactions between components called \textit{interaction scenarios} in Fig.~\ref{figure:complex_interaction}.
Each interaction scenario is related to some specific API and associated with arrows of a particular style.
For instance, the device driver provides initialization and exit functions (long-dashed line), device-specific callbacks (dotted line), interrupt handlers (chained line) and, in turn, it calls library functions from the bus driver (thin line) and the kernel (short dashed line).
We refer to corresponding function calls and read or write accesses to global variables as \textit{interaction events}.
Thus, each interaction scenario can be considered as a set of feasible sequences of interaction events that can happen during real program execution.
We assume that all interaction events in an interaction scenario should happen in the context of the same thread or process, but each event happening in the environment is followed by the execution of some code from the program fragment.

The environment model might include different models of interaction scenarios, depending on the decomposition of a program into fragments.
It is possible to verify either a device driver separately from other components, both device and bus drivers together, or even add to them extra files from the operating system kernel.

The next subsection describes the semantics of the used representation of environment models and an approach to their generation.

\subsubsection{Intermediate Model}

We propose a notation for the so-called \textit{intermediate model} to specify environment models of interaction scenarios, or just \textit{scenario models} for brevity.
In simple words, the intermediate model describes events that can happen in the environment of a program fragment in a structured way.
The language definition is out of the paper's scope, it is given in the Klever documentation.
In this section, we present the semantics of the notation.

An intermediate model aims at modeling scenarios before their translating to a C source code of the environment model.
Models in the notation can be either generated or developed by a user.

Let the interface of a program fragment be $I = <V_p, F_p, R_p, T_p>$.
$V_p$ is a set of global variables, $F_p$ is a set of function declarations, $R_p$ is a set of macros, and $T_p$ is a set of type definitions provided in the C programming language.
The "p" suffix means that these objects are declared or defined in the program fragment.

An intermediate model for the program fragment with interface $I$ consists of the following elements:
\[ M_I = <V_e, F_e, R_e, T_e, E_F, E_T > \]
\[ E_T = \varepsilon_1 \parallel .. \parallel \varepsilon_n \]

Where $V_e, F_e, R_e, T_e$ are supplementary global variables, functions, macros and type definitions provided in the C programming language as separate C source files or headers.
The "e" suffix helps to distinguish environment-related objects from the program fragment's ones.
$E_F$ and $E_T$ represent external functions and threads spawned by an environment correspondingly.

The set of function definitions $F_e$ contains two kinds of functions (auxiliary and project-specific ones). Auxiliary functions are reused between projects, and many of them depend on a used verification tool.
Models of POSIX or C standard library functions belong to project-specific functions.

$E_F$ and $E_T$ denote definitions of models of external functions and specific separate threads for events happening in the environment correspondingly.
Models of external functions specify undefined functions from $F_p \setminus F_e$ (functions that don't have definitions or models in either the program fragment or environment model supplementary files).
Both the environment thread and external function models specify scenario models.
The former describes events executed in a separate thread spawned by an environment.
The latter specifies sequences of interaction events executed in an already existing thread.
The notation limits an application scope to parallel programs with shared memory.
The semantics of environment thread models assume that interaction events would happen in threads started simultaneously in the environment model\footnote{However, the word "run" here is used for convenience because the environment model and the program fragment are never intended for real execution but only for verification.}.

A scenario model $\varepsilon$ can be considered as a transition system:
\[ \varepsilon = <\mathcal{V}, \mathcal{A}, \mathcal{R}, \alpha_0>\]

Where $\mathcal{V}$ is a set of \textit{labels}, $\mathcal{A}$ is a set of actions, $\mathcal{R}$ is a transition relation of the scenario model.
Labels are local variables of scenario models in simple words.
States of the transition system correspond to different assignments of values to labels from $\mathcal{V}$ and variables $V_p \cup V_e$ (global variables defined in the program fragment and environment model respectively).
Actions describe state transitions and interaction events.
Each label has a type defined in the program fragment or the standard C library.
Each scenario model has the first action denoted as $\alpha_0$.
There are three kinds of actions: sending/receiving a signal, base block and jump.

Base blocks contain the C code, such as entry point calls or accesses to the heap memory and global variables.
Each base block action is a couple:
\[\alpha = <\varphi, \beta> \]
where $\varphi$ is a precondition defined as a C logical expression over variables and labels from $\mathcal{V} \cup V_p \cup V_e$.
$\beta$ contains C statements over variables and labels from $\mathcal{V} \cup V_p \cup V_e$ and functions from $(F_p \cup F_e) \setminus E_F$.
The base block code should follow the C programming language syntax.
Each block should have a single control entry and exit points without any \textit{goto} statements or incomplete switches, loops or conditional operators.
It is also forbidden to introduce new variables in base blocks.

Signal exchanges allow for describing data and ordering dependencies between actions in different scenario models.
Signals behave according to the Rendezvous synchronization model proposed by Tony Hoare~\cite{Hoare:1978:CSP}.

Let us consider a scenario model $\varepsilon_i = <\mathcal{V}_i, \mathcal{A}_i, \mathcal{R}_i, \alpha_{0_i}>$ that sends a signal to $\varepsilon_j = <\mathcal{V}_j, \mathcal{A}_j, \mathcal{R}_j, \alpha_{0_j}>$.
Sending and receiving actions are $\alpha_i \in \mathcal{A}_i$ and $\alpha_j \in \mathcal{A}_j$:
\begin{align*}
&\alpha_i = <\varphi_i, \pi_i, l_i>
&\alpha_j = <\varphi_j, \pi_j, l_j, \psi_j>
\end{align*}
Constants $l_i$ and $l_j$ are signal names.
Signal exchange happens if and only if $l_i = l_j$.
C logical expressions $\varphi_i$, $\varphi_j$ and $\psi_j$ over $\mathcal{V}_j \cup V_p \cup V_e$ define the precondition and postcondition.
Sending action $\alpha_i$ does not have a postcondition because its local variables stay unchanged.
Two vectors of labels $\pi_i: {v_1, ..., v_k}$ where $v_t \in \mathcal{V}_i$ for $t \in 1..k$ and $\pi_j: {u_1, ..., u_k}$ where $u_t \in \mathcal{V}_j$ for $t \in 1..k$ describe data transfer: $\forall t = 0..k:~u_t := v_t$.
Types of corresponding labels at the same positions in vectors should match each other.

Jumps are actions that help to implement loops and recursion.
An entrance to a jump replaces the transfer relation with new transfer relation rules.

The order of actions ($\mathcal{R}$ and $\alpha_0$) in the example is specified in the \textit{process} entry using a simple language.
There is a notation for the following kinds of actions:
\begin{itemize}
    \item $<name>$ is a base block action;
    \item $(name)$ is a signal receiving action. Where $(!name)$ means that the scenario model waits for a signal to start.
    \item $[name]$ is a signal sending action.
    \item ${jump}$ is a jump that just specifies a new sequence of actions to do. Each jump action has its \textit{process} entry.
\end{itemize}

An order of actions is described using two operators:
\begin{itemize}
    \item $.$ is a sequential combination operator.
    \item $|$ is a non-deterministic choice operator.
\end{itemize}

One can use parentheses in expressions; a sequential combination operator has a higher priority.

\begin{figure}
  \centering
    \includegraphics[width=0.5\textwidth]{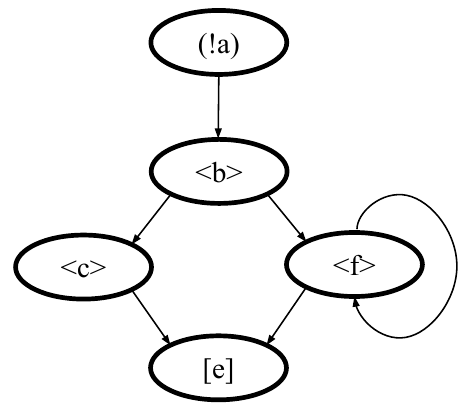}
    \caption{Example of an interaction scenario transfer relation.}
    \label{figure:process_order}
\end{figure}

There is an example of a transfer relation of a scenario model illustrated in Fig.~\ref{figure:process_order}.
The order can be defined by the following expressions:
$(!a).<b>.(<c> | {d}).[e]$ and a jump action ${d} = <f>.{d} | [e]$.

The notation of intermediate models is sufficient for specifying realistic environment models, as it allows combining an event-oriented description of scenarios with fragments in C.
It gives more freedom to describe the event-driven environment model, dependencies between events, and a  non-deterministic behavior.
Descriptions of actions are separated from definitions of signals and transfer relations of scenario models.
Such organization simplifies the comprehension of the whole model structure.
It is easy to illustrate it as a graph.
The model-generating process allows weaving the environment model into files of program fragments without touching the source code.
It simplifies the work of dealing with different versions and configurations of a program.

The drawbacks of the approach are the large size of JSON files and the redundancy of many features if an environment model is simple and requires mostly function models.
In this case, it is easier to provide the model directly in C. This case is also supported in Klever.

A simplified example of an intermediate model and its description can be found in Appendix~\ref{Appendix:Env}.

\subsubsection{Generating Environment Models}

\begin{figure}
\centering
\includegraphics[scale=0.6]{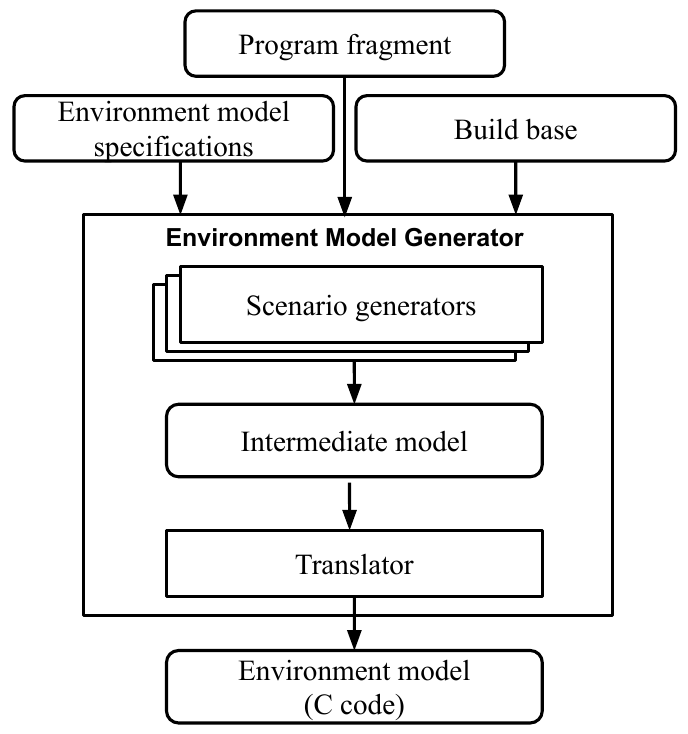}
\caption{Environment Model Generator workflow.}
\label{figure:em}
\end{figure}

Environment Model Generator (EMG) performs the synthesis of environment models for program fragments according to the workflow illustrated in Fig.~\ref{figure:em}.
As an input, EMG receives a program fragment, build base, and optional environment model specifications developed manually.
As its output, EMG synthesizes an environment model in the aspect-oriented extension of the C programming language.
CIF weaves the program fragment source code to add the environment model, but this step follows EMG's work\footnote{The next subsection and Appendix~\ref{Appendix:CIF} contain more details about the aspect-oriented extension of the C programming language and CIF.}.

The EMG's synthesis process consists of two steps:
\begin{enumerate}
    \item First, \textit{scenario generators} independently synthesize scenario models.
    Their combination forms an intermediate model.
    \item Then, the \textit{translator} generates the C code of the environment model from the provided intermediate model.
\end{enumerate}

Each scenario generator provides scenario models that can communicate over signals.
Scenario generators are parts of program-specific adaptations and can implement various implementations.
The open-source implementation contains generators for the Linux device drivers, while the implementation just combines manual environment model specifications.
Notations of environment model specifications for various scenario generators may differ.
It is convenient to use a pipeline of several scenario generators for different kinds of scenario models in practice.
For large event-oriented programs, the development of scenario generators may take a considerable time.

In some cases, it is convenient to manually adjust or develop an intermediate model or its part.
We propose to use a specific scenario generator to enable that.

The translator generates the main single C source file with the entry point and several files in the aspect-oriented extension of the C programming language called \textit{aspect files}.
The first EMG translation step is the extraction of the C code denoted as $V_e, F_e, R_e, T_e$ from the provided intermediate model.
This part of the model does not need any translation and can be added to the corresponding files.

Then the translator should generate a function for each scenario model.
We refer to such functions as \textit{control functions}.
We give an example of a control function in Appendix~\ref{Appendix:Env}.
To generate a control function, the translator creates a control function with local variables per each label from $\mathcal{V}$, adds C control operators to implement the order of interaction events defined by the intermediate model, and adds an implementation per event, which is explained below.
Subsequently, the translator determines how to replace references to labels in actions with the newly generated control function local variables.



There is a sound approach for translating signal exchanges to C code without losing scenarios.
Unfortunately, preliminary experiments indicated that the existing verification tools could not cope with complicated parallel models when checking memory safety or reachability properties.
Thus, we have implemented a simplified translator that implies additional restrictions on models and misses some possible scenarios.
It allows only scenario models that can receive signals as the first and the last actions.
The resulting code is sequential, and it calls corresponding control functions in place of signal dispatches to scenario models.
Such optimization makes the resulting model straightforward but does not capture the interleaving of actions from different scenario models that may run simultaneously.

There is a specific translating algorithm for getting simplified parallel models to check data races described in~\cite{Andrianov:2020:ACS}.
The obtained model is parallel, but it still does not allow signal exchange between the first and the last actions.



The translator generates a control flow graph that determines the order of actions using \textit{if} and \textit{switch} operators instead of "|" and ".", and \textit{goto} jumps instead of auxiliary \textit{jump} actions to proceed with the generating control functions.
After the control flow graph is ready, the translator supplies it with the code of base block actions and newly generated code for receiving and sending actions in particular places.

CIF also adds calls to control functions from aspect files to original source files of the program to trigger the execution of control functions at points where functions that correspond to function scenario models are called.

Some extra details about environment modeling in the Klever framework can be found in~\cite{Khoroshilov:2015:MES,Novikov:2018:VOS,Zakharov:2018:CEM} and in the user documentation\footnote{https://klever.readthedocs.io/en/latest/dev\_env\_model\_specs.html}.

\subsubsection{Modeling Complicated and Undefined Functions}

As we already mentioned, it may be necessary to model complicated and undefined functions that influence the verification results.
There is an example of a model of the Linux kernel memory allocation function \textit{kmalloc()} in Fig.~\ref{listing:memmodel}.
Fig.~\ref{listing:memmodelAspect} shows how this model is bound with the original function from the Linux kernel.

The model reduces the \textit{ldv\_kmalloc()} function to \textit{malloc()} which is known by any verification tool that participates in \mbox{SV-COMP}.
The model takes into account that \textit{malloc()} does not return NULL according to the \mbox{SV-COMP} rules and that in the Linux kernel there are reserved error codes that cannot be valid pointer addresses (they are filtered out by \textit{ldv\_is\_err()}).

\begin{figure}
\centering
\begin{minted}[frame=single,fontsize=\scriptsize]{C}
/* Definitions of types like gfp_t. */
#include <linux/types.h>
/* Definition of ldv_assume() that shrinks paths with a false value of its argument. */
#include <ldv/verifier/common.h>
/* Definition of ldv_undef_int() that returns any integer type value. */
#include <ldv/verifier/nondet.h>

void *ldv_kmalloc(size_t size, gfp_t flags) {
    void *res;

    /* This allows to check for flags within requirement specifications. */
    ldv_check_alloc_flags(flags);

    /* Returns either NULL or a pointer to a valid kernel memory nondeterministically. */
    if (ldv_undef_int()) {
        /* Always successful according to SV-COMP rules. */
        res = malloc(size);
        ldv_assume(!ldv_is_err(res));
        return res;
    } else
        return NULL;
}

long ldv_is_err(const void *ptr) {
    return ((unsigned long)ptr > (unsigned long)-MAX_ERRNO);
}
\end{minted}
\caption{Model of a memory allocating function.}
\label{listing:memmodel}
\end{figure}

\begin{figure}
\centering
\begin{minted}[frame=single,fontsize=\scriptsize]{C}
around: execution(static inline void *kmalloc(size_t size, gfp_t flags))
{
    return ldv_kmalloc(size, flags);
}
\end{minted}
\caption{Aspect file replacing the \textit{kmalloc()} function definition with its model.}
\label{listing:memmodelAspect}
\end{figure}

Using \textit{ldv\_undef\_int()} causes verification tools to traverse both paths when analyzing the target program with the given model.
The first path corresponds to successful memory allocation, while the second one is a failure.
This allows us to reveal possible faults in error handling paths that are unlikely to happen during normal operation and testing.

\textit{ldv\_kmalloc()} invokes \textit{ldv\_check\_alloc\_flags()} that may be defined by requirement specifications to check for memory allocation function flags.
For instance, one can ensure that the flag \textit{GFP\_ATOMIC} is passed when allocating memory in the interrupt context and when holding spinlocks within the Linux kernel.

Fig.~\ref{listing:changedAPIAspect} gives an example of support for different versions of the Linux kernel.
In this example, the same model \textit{ldv\_is\_err()} is used since only the return type of \textit{IS\_ERR()} changed, but its return values remained the same.
In the case of changes in the semantics of the API, one may need to introduce different models.

\begin{figure}
\centering
\begin{minted}[frame=single,fontsize=\scriptsize]{C}
around: execution(static inline long IS_ERR(const void *ptr))
{
	return ldv_is_err(ptr);
}

/* Starting from Linux 3.15 IS_ERR() returns Boolean values. */
around: execution(static inline bool IS_ERR(const void *ptr))
{
	long ret;
	ret = ldv_is_err(ptr);
	return (bool)ret;
}
\end{minted}
\caption{Aspect file taking into account changes in the Linux kernel API.}
\label{listing:changedAPIAspect}
\end{figure}

One can find additional information on this topic in the user documentation\footnote{https://klever.readthedocs.io/en/latest/dev\_common\_api\_models.html}.

\subsection{Specifications of Requirements}

For checking requirements that do not correspond explicitly to any of the properties supported by verification tools~\cite{Beyer:2021:SVC}, we suggest to weave an additional source code into a program.
This extra source code should express the requirements using one of the supported properties.
For instance, rules of correct usage of a particular API can be formulated as the unreachability of the error function, like in Fig.~\ref{listing:originalSourceCode} and Fig.~\ref{listing:wovenInSourceCode}.
For some requirements, it may be hard or even impossible to express them using additional C expressions and statements.
In this case, one has to leverage specific means of verification tools, e.g., this may be the case for finding data races~\cite{Andrianov:2020:ACS}.

\begin{figure}
\centering
\begin{minted}[frame=single,fontsize=\scriptsize]{C}
/* A device driver can register just one device at a time. */
int register_device(void) {
    ...;
}
/* A device driver can unregister a device just after it registers it. */
void unregister_device(void) {
    ...;
}
\end{minted}
\caption{Original program source code.}
\label{listing:originalSourceCode}
\end{figure}

\begin{figure}
\centering
\begin{minted}[frame=single,fontsize=\scriptsize]{C}
bool is_device_registered = false;

int register_device(void) {
    if (is_device_registered)
        __VERIFIER_error();
    is_device_registered = true;
    ...;
}
void unregister_device(void) {
    if (!is_device_registered)
        __VERIFIER_error();
    is_device_registered = false;
    ...;
}
\end{minted}
\caption{Modified program source code.}
\label{listing:wovenInSourceCode}
\end{figure}

If one expresses weakly related requirements using the same property, it is possible to check them simultaneously, but we do not recommend this due to the following issues.
The first reason is that it is not an easy task for verification tools to distribute available computational resources fairly between various requirements.
The second reason is that most verification tools stop after they find a first violation of a checked property.
Therefore, detecting a first fault or a false alarm can prevent finding other faults.

Below, we provide an example of a requirements specification.
Appendix~\ref{Appendix:ReqSpecBase} contains its high-level description.
If the reader is not interested in this, he/she can proceed to the next subsection.

Fig.~\ref{listing:modulemodel} and Fig.~\ref{listing:moduleModelAspect} contain an example of a requirements specification that reduces checking of correct usage of the module reference counter API in the Linux kernel to a reachability problem.
The requirements specification introduces a model state represented by the global variable \textit{ldv\_module\_refcounter} initialized by 0.
The model state is changed within model functions \textit{ldv\_module\_get()}, \textit{ldv\_try\_module\_get()}, and \textit{ldv\_module\_put()} according to the semantics of the corresponding API.
These model functions are bound with the original ones by the aspect file shown in Fig.~\ref{listing:moduleModelAspect}.

\begin{figure}
\centering
\begin{minted}[frame=single,fontsize=\scriptsize]{C}
/* Definition of ldv_assert() that calls __VERIFIER_error() or dereferences NULL. */
#include <ldv/verifier/common.h>
#include <ldv/verifier/nondet.h>

/* NOTE Initialize module reference counter at the beginning */
static int ldv_module_refcounter = 0;

void ldv_module_get(struct module *module)
{
    if (module)
        /* NOTE Increment module reference counter */
        ldv_module_refcounter++;
}

int ldv_try_module_get(struct module *module)
{
    if (module && ldv_undef_int()) {
        /* NOTE Increment module reference counter */
        ldv_module_refcounter++;
        /* NOTE Successfully increment module reference counter */
        return 1;
    }

    /* NOTE Could not increment module reference counter */
    return 0;
}

void ldv_module_put(struct module *module)
{
    if (module) {
        if (ldv_module_refcounter == 1)
            /* ASSERT Decremented module reference counter should be greater than its
                      initial state */
            ldv_assert();
        /* NOTE Decrement module reference counter */
        ldv_module_refcounter--;
    }
}

void ldv_check_final_state(void)
{
    if (ldv_module_refcounter != 0)
        /* ASSERT Module reference counter should be decremented to its initial value
                  before finishing operation */
        ldv_assert();
}
\end{minted}
\caption{Example of expressing requirements as a reachability problem.}
\label{listing:modulemodel}
\end{figure}

\begin{figure}
\centering
\begin{minted}[frame=single,fontsize=\scriptsize]{C}
around: call(void __module_get(struct module *module) {
	ldv_module_get(module);
}

around: call(bool try_module_get(struct module *module)) {
	return ldv_try_module_get(module);
}

around: call(void module_put(struct module *module)) {
	ldv_module_put(module);
}
\end{minted}
\caption{Aspect file replacing kernel function calls with model ones.}
\label{listing:moduleModelAspect}
\end{figure}

The considered requirements specification makes 2 checks by \textit{ldv\_assert()}.
The first one is within \textit{ldv\_module\_put()}.
It is intended to find out when Linux kernel loadable modules decrement the module reference counter without first incrementing it.
The second check is within \textit{ldv\_check\_final\_state()}.
It tracks that modules should decrement the module reference counter to its initial value before finishing their operation.

To emphasize statements that are most relevant to checked requirements, we suggest using special comments.
Such comments in the example begin with the keywords \textit{NOTE} and \textit{ASSERT}.
These comments are used during preliminary processing and visualization of violation witnesses considered in the following subsections.


One can find more details about the development of requirement specifications in the corresponding section of the user documentation\footnote{https://klever.readthedocs.io/en/latest/dev\_req\_specs.html}.

\subsection{Verification Task Generation}

Klever generates a verification task for each pair of a program fragment and a requirements specification.
Below, we consider particular steps of this process.

\subsubsection{Weaving and Merging of Source Files and Models}

First, CIF\footnote{One can find more details about CIF and CIL in Appendix~\ref{Appendix:CIF} and Appendix~\ref{Appendix:CIL} respectively.} weaves the source code of the program fragment with aspect files from the requirements specification and the environment model.
At this step, preprocessing is also performed.
Thereafter, there are several instrumented and preprocessed original C source files as well as additional preprocessed model C source files.
CIL combines all the source files.
After all, there is a single preprocessed C file \textit{cil.i} prepared for an immediate run of a verification tool.

\subsubsection{Configuration of Automatic Software Verification Tools}

It may be helpful to experiment with different configurations of verification tools in practice.
There is also a need to specify different sets of verification tool options for various requirement specifications.
Such sets may even differ for various versions of the same tool.

To simplify the configuration process for end-users, we propose to prepare several so-called \textit{verifier profiles} in advance.
Each requirements specification refers to one of verifier profiles by default.
In addition, a user can choose any other available variant without investing time into learning particular verification tool options and capabilities.
There is a verifier profiles example in Appendix~\ref{Appendix:VP}.


At the moment, Klever supports CPAchecker~\cite{Beyer:2011:CTC} as a verification tool backend and there is a bunch of verifier profiles for it.
To integrate new verification tools within Klever, users need to do at least the following: 
\begin{itemize}
\item Describe specific verification tool options suitable for checking requirement specifications as extra verifier profiles.
\item Provide verification tool binaries to a Klever deployment script.
\end{itemize}

One can learn more about configuration of verification tools in the corresponding section of the user documentation\footnote{https://klever.readthedocs.io/en/latest/dev\_verifier\_profiles.html}.

\subsubsection{Final Preparation of Verification Tasks}

At the final step, Klever combines everything obtained thus far into verification tasks.
Each verification task consists of the following set of files:
\begin{enumerate}
\item \textit{cil.i} (its generation was described above).
\item A property file, like in Fig.~\ref{listing:property}.
Its content is provided by a verifier profile referred to by a checked requirements specification.
One can see a complete list of properties supported by various verification tools in~\cite{Beyer:2021:SVC}.
\item A task definition file\footnote{https://github.com/sosy-lab/benchexec/blob/main/doc/task-definition-example.yml}.
Klever uses it only to bind two previous files,
like in Fig.~\ref{listing:task}.
\item A benchmark definition file\footnote{https://github.com/sosy-lab/benchexec/blob/main/doc/benchexec.md\#input-for-benchexec}.
It connects all previous files as well as specifies various verification tool options provided by a user in various configuration files.
An example of this file is shown in Fig.~\ref{listing:benchmark}.
\end{enumerate}

\begin{figure}
\centering
\begin{minted}[frame=single,fontsize=\scriptsize]{ini}
CHECK( init(main()), LTL(G ! call(__VERIFIER_error())) )
\end{minted}
\caption{Property file example.}
\label{listing:property}
\end{figure}

\begin{figure}
\centering
\begin{minted}[frame=single,fontsize=\scriptsize]{yaml}
format_version: '1.0'

input_files: 'cil.i'

properties:
  - property_file: safe-prps.prp
\end{minted}
\caption{Task definition example.}
\label{listing:task}
\end{figure}

\begin{figure}
\centering
\begin{minted}[frame=single,fontsize=\scriptsize]{xml}
<?xml version="1.0" ?>
<benchmark hardtimelimit="300" timelimit="270" tool="cpachecker">
    <rundefinition>
        <option name="-setprop">counterexample.export.exportExtendedWitness=true</option>
        <option name="-ldv-bam"/>
        ...
    </rundefinition>
    <tasks>
        <include>cil.yml</include>
    </tasks>
    <propertyfile>safe-prps.prp</propertyfile>
</benchmark>
\end{minted}
\caption{Benchmark definition example.}
\label{listing:benchmark}
\end{figure}

\subsection{Scheduling and Monitoring}


Klever~Scheduler operates with two types of objects: verification tasks and verification jobs.
A \textit{verification job} is a set of files including a build base\footnote{As a rule, build bases are not explicitly included in verification jobs. They are located and found in the Klever deployment directory.}, specifications, and configuration files that are necessary to start automatic verification.
Verification jobs are solved by a specific Klever component named Klever~Core.
During the solution of verification jobs, it generates verification tasks according to the approaches described above.
Each Klever~Core instance can generate verification tasks in parallel to speed up the entire verification process.

Monitoring of available computational resources and their fair distribution between verification jobs and verification tasks are responsibilities of Klever~Scheduler.
Klever Scheduler respects verification task priorities specified by users.
Also, Klever~Scheduler supports canceling verification jobs.
Some verification tasks can require considerably lesser computational resources than specified by the user.
To avoid useless memory reservations, we perform speculative scheduling, trying to run a verification tool with lesser limitations first.

We use BenchExec\footnote{See Appendix~\ref{Appendix:BenchExec} for more information.} to isolate runs of Klever~Core and verification tools as well as to measure and limit computational resources consumed by them at a single machine.
It also lets you get verification results from different verification tools in a unified format.

Klever~Scheduler currently supports parallel solution of verification jobs and tasks on a single machine.
Also, it can solve verification tasks in parallel using VerifierCloud\footnote{https://vcloud.sosy-lab.org}.


\subsection{Preliminary Processing of Verification Results}

One of the most important functions of the framework is the presentation of verification results to users.
The Klever component named Verification Result Processor (VRP) performs preliminary processing of verification results for each solved verification task.

Violation witnesses generated by verification tools are machine-readable, and a user can hardly use them in their raw format.
Moreover, verification tools can miss some details and even parts of corresponding error paths.
We supported an extended witness format for CPAchecker to have all the necessary pieces of information\footnote{https://klever.readthedocs.io/en/latest/dev.html\#extended-violation-witness-format}.
In addition, it allows to add some important internal information from verification tools to the violation witness.
For instance, places where leaked memory is allocated can be specially noted during memory-safety checking.
Other verification tools do not provide violation witnesses using our format; thus, their visualization may not be as good as the CPAchecker ones.

VRP translates violation witnesses into a Klever internal format that is more convenient for visualization and assessment purposes\footnote{https://klever.readthedocs.io/en/latest/dev.html\#error-trace-format}.
Using information from violation witnesses, it finds corresponding expressions and statements within \textit{cil.i} to be presented to users and adds references to corresponding lines of source files.
Also, at this stage, VRP parses special comments from environment models and requirement specifications.
This helps to distinguish important pieces of information like invocations of program fragment entry points and changes to model states.
In contrast, VRP marks all parts of violation witnesses that seem irrelevant for found violations of checked requirements so that they can be hidden during subsequent visualization.

The CPAchecker verification tool can provide code coverage reports in addition to witnesses~\cite{Castano:2017:MCE}.
These reports are in the GCC test coverage format (GCOV).
For its visualization, one can use standard tools like LCOV\footnote{http://ltp.sourceforge.net/coverage/lcov.php}.
Code coverage reports are an important artifact to establish verification in practice.
They reflect parts of the program such as lines of code, branches, and functions that are actually verified.
This information is essential for estimating the quality of an environment model, since neither violation nor correctness witnesses provide data on actual program paths.
Code coverage reports help to understand which program entry points should be invoked additionally by environment models.

VRP converts code coverage reports to an internal format to facilitate their visualization\footnote{https://klever.readthedocs.io/en/latest/dev.html\#code-coverage-format}.
Like for violation witnesses, it adds references to the original source files.
Also, it merges reports issued for individual verification tasks to get a single code coverage report for the program for each requirements specification.
At this step, VRP calculates statistics for each subdirectory of the program source tree, considering either source files considered by a verification tool or all source files from a build base.
These sets of source files can differ since the verification tool may miss analysis for certain source files, e.g., due to some internal failures or if some source files are not included by a considered program configuration.

\subsection{User Interface}

The Klever software verification framework implements a user interface as a web-server named Klever~Bridge.
Below, we consider two major use cases of Klever~Bridge.

\subsubsection{Managing Verification Processes}

\begin{figure}
\includegraphics[width=1\textwidth]{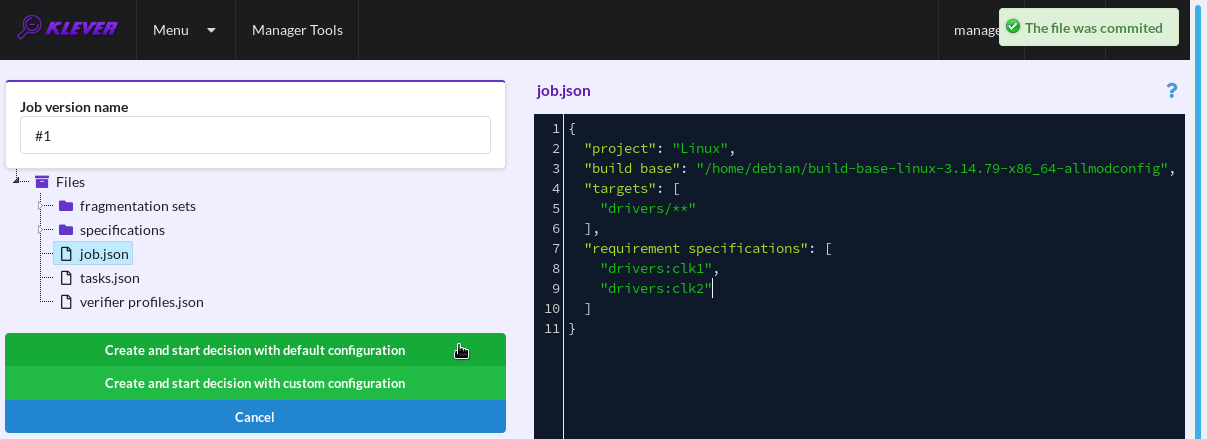}
\caption{Creation of a new verification job.}
\label{fig:jobCreation}
\end{figure}

To obtain verification results, users should prepare all the necessary data and then start verification.
At this stage, Klever Bridge provides the following facilities:

\begin{itemize}
\item Each new verification activity starts with the creation of a new verification job (Fig.~\ref{fig:jobCreation}).
Verification jobs consist of specifications and configuration files that are necessary to start the verification process and have attributes such as a name, date of creation, author, access rights, etc.
Verification jobs can be created based on templates for corresponding project-specific adaptations or existing verification jobs.
To proceed to verification, users should choose program fragments and requirements to be checked.
Besides, to get better verification results for particular programs and requirements, we suggest incrementally tuning various configuration options and improving specifications.
Klever~Bridge provides a featured text editor to complete these tasks.
\item Klever~Scheduler starts the execution of Klever~Core as soon as the user wants to solve his/her newly created verification job and there are available computational resources.
Klever~Core reports verification results when they become available.
It is worth noting that since verification can take a considerable time, Klever~Bridge presents verification results to experts as soon as they appear.
In this case, they can proceed to their analysis faster; in particular, it is possible to understand that something was done wrong without waiting for all verification results.
\item Klever~Bridge provides information on running and already solved verification jobs.
For running verification processes, it presents their progress: the number of already solved and the total number of verification tasks, elapsed time, and approximate remaining time.
The latter is evaluated by Klever~Core and Klever~Scheduler based on accumulated statistics.
\end{itemize}

\subsubsection{Expert Assessment of Verification Results}

\begin{figure}
\includegraphics[width=1\textwidth]{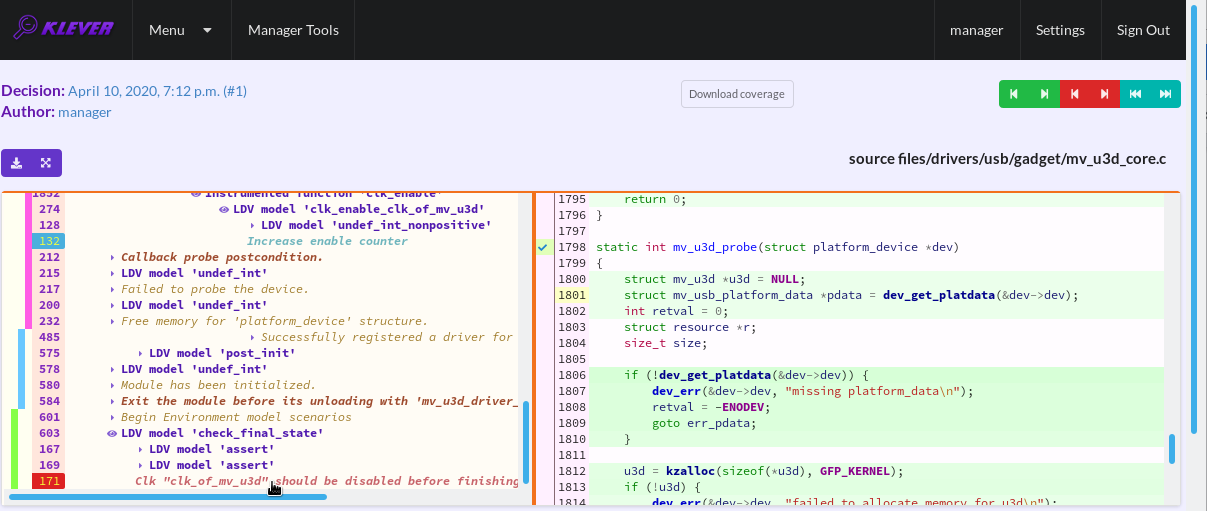}
\caption{Visualized violation witness and code coverage.}
\label{fig:violationWitness}
\end{figure}

\begin{figure}
\includegraphics[width=1\textwidth]{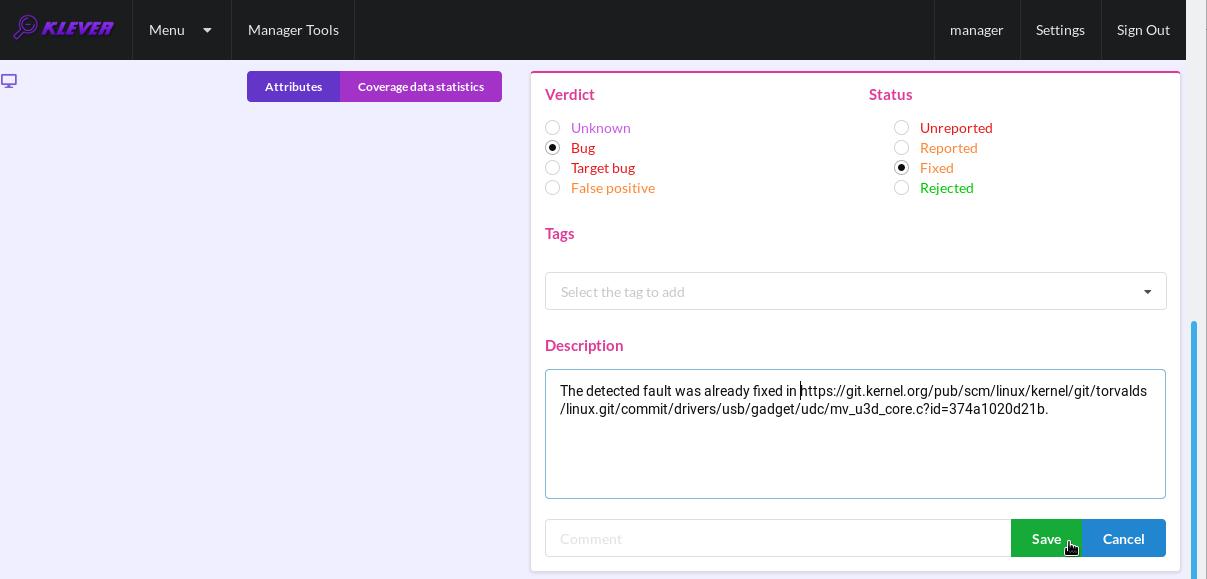}
\caption{Creation of a new expert mark.}
\label{fig:markCreation}
\end{figure}

To simplify the analysis of verification results by experts, we implemented the following facilities in Klever~Bridge:
\begin{itemize}
\item Visualization of violation witnesses, code coverage and internal failures (Fig.~\ref{fig:violationWitness}).
The primary goal of this visualization is to hide from experts as many irrelevant, according to domain knowledge, details as possible.
Besides, all verification results are closely related to the original source files.
\item Evaluation of verification results by creating expert marks (Fig.~\ref{fig:markCreation}).
Expert marks can be associated with verification results, either manually or automatically.
The automatic assessment saves a great deal of time by avoiding the analysis of similar verification results when different versions or configurations of the same program are verified.
Often, in these cases, violation witnesses and internal failure reasons do not differ considerably, and Klever~Bridge matches them by pretty simple rules.
Experts can provide each mark with a detailed description and tags.
To further simplify the analysis, Klever~Bridge keeps all the history of mark changes.
\item Showing various statistics over verification results that can help to understand a general picture.
For instance, it can be very useful to see how many warnings were yielded for a particular verification job, what warnings correspond to faults and false alarms, what are the most significant reasons of false alarms, and so on.
\item Users can compare both verification jobs as sets of files and their verification results.
This is useful both for tracking changes for various versions and configurations of target programs and for estimating changes during the development of project-specific adaptations.
\end{itemize}
It might be useful to integrate expert marks with issue tracking systems.
However, it might be worth noting that a mark does not always correspond to a real bug.
Marks might help to distinguish false alarms or issues in the environment models.

One can find more examples and detailed explanations of some actions with the Klever user interface in the tutorial\footnote{https://klever.readthedocs.io/en/latest/tutorial.html}.

\section{Project-Specific Adaptations}
\label{section:adaptations}

This section describes publicly available project-specific adaptations delivered together with the Klever software verification framework.
The application domain of Klever is, however, not limited to the programs described in this section.

There are two usable adaptations --- for the Linux kernel and BusyBox.
The adaptation for the Linux kernel demonstrates capabilities of verification of complex event-driven software.
The Busybox adaptation is a proof of concept of how Klever is applicable to user-space programs.


\subsection{Adaptation for the Linux Kernel}

The Linux kernel architecture is monolithic --- it is loaded into RAM during operating system startup, and then it operates completely in the same shared address space.
The Linux kernel implements main operating system functionalities such as scheduling, memory management, interprocess communication, interrupt handling, a network stack, and so on.
Besides, it supports modules that can extend its functionality further.
As a rule, device drivers, file systems, network protocols, and audio codecs are developed as modules.
Modules can be either statically compiled into the kernel or dynamically loaded, but in both cases, they become parts of the monolithic structure.

We consider the whole Linux kernel, except loadable modules, as a set of subsystems.
Each subsystem usually consists of all the C source files inside a specific subdirectory of the kernel source tree.
Typical personal computers can load several hundred modules, but that is a small fraction of all available modules.
There are more than \SI{8000} modules in Linux 5.12-rc3 compiled for the \textit{allmodconfig} configuration.
The average module size is about 1.7 KLOC with a median of 500 lines.
The average subsystem size is 7 KLOC with a median of 2 KLOC.

The adaptation for the Linux kernel includes:
\begin{itemize}
    \item Tactics for decomposition and composition.
    They extract modules and subsystems from the Linux source code and combine them together.
    \item A pipeline of several scenario generators to obtain environment models.
    \item A relatively large set of environment model specifications.
    \item A set of requirement specifications allowing one to check memory safety, correct usage of the Linux kernel API, and the absence of data races.
\end{itemize}

Below, these items are treated in detail.
In the given section, we do not provide verification results for the Linux kernel, since it is a matter for the following section.

\subsubsection{Decomposition and Composition Tactics}

There is only one tactic implemented for decomposing the Linux kernel source code.
This tactic uses a build command graph from the Linux kernel build base as an input.
It iterates over LD (linker) and AR (archiver) build commands.
The tactic divides them into two types, depending on the extension of the output files.
Files with the extension ".ko" correspond to loadable kernel modules.
Subsystems are associated with files "built-in.o" or "built-in.a" in recent versions of the Linux kernel.

The decomposition tactic generates a separate program fragment for each selected build command.
It traverses the build command graph to extract CC (compiler) build commands whose output files become a part of the module or subsystem.
Each CC build command has an input C source file, and such files together form the final program fragment.

The composition step is optional, so we do not consider it here in detail.
There are two composition strategies that allow to combine modules or subsystems that call exported functions from each other using a greedy algorithm.

\subsubsection{Environment Modeling}

A program fragment can contain a single module or several modules and subsystems.
An interface for the program fragment can include:
\begin{itemize}
    \item \textit{Subsystem initialization functions}.
    These functions are entry points of subsystems, and the Linux kernel calls them in a specific order at boot time.
    \item \textit{Module initialization and exit functions}.
    Pairs of such functions are entry points for modules.
    The Linux kernel calls them when loading and unloading modules from memory, respectively.
    \item \textit{Callbacks}.
    Each separate group of callbacks implements operations for handling events relevant to a particular resource or device.
    Callbacks are registered and deregistered in the Linux kernel by calling special functions.
    \item \textit{Exported functions}.
    Any module or subsystem can implement functions that are exported and used by other Linux kernel components.
\end{itemize}

\begin{figure}
    \centering
    \includegraphics[scale=0.5]{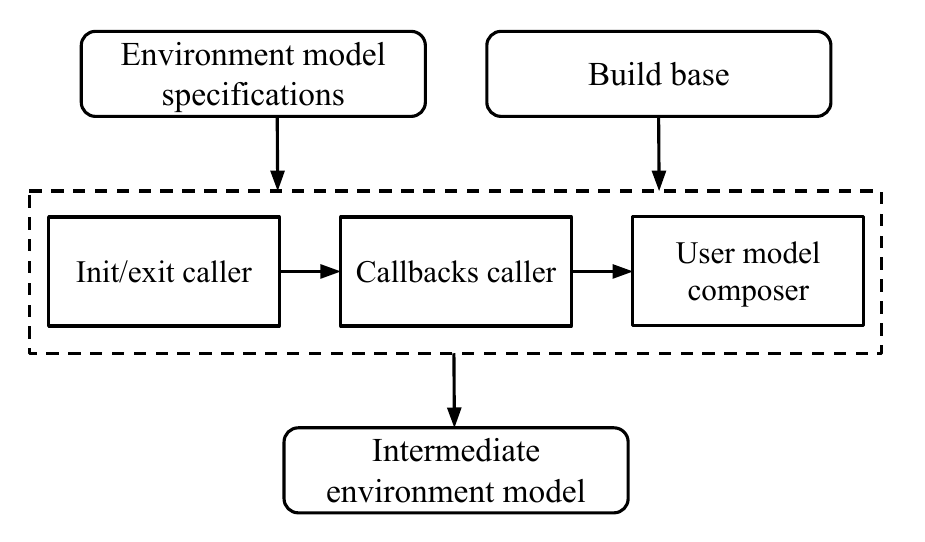}
    \caption{Pipeline of scenario generators.}
    \label{fig:emgpipeline}
\end{figure}

There is a pipeline of three scenario generators to compose environment models according to Fig.~\ref{fig:emgpipeline}:
\begin{enumerate}
    \item \textit{Init/exit caller}. It is a scenarios generator for calling subsystem initialization functions and model initialization and exit functions.
    \item \textit{Callbacks caller}. The scenarios generator invokes Linux kernel callbacks.
    \item \textit{User model composer}. The scenarios generator adds handcrafted parts to the intermediate model.
\end{enumerate}

\textit{Init/exit caller} accepts as input a specification with an order for calling the subsystem's initializing functions.
Klever contains such a specification that enumerates a list of relevant macros for this purpose.
The scenarios generator checks if these macros and \textit{module\_init}/\textit{module\_exit} macros are used in the source code of the program fragment.
If so, it groups modules' initialization and exit functions in pairs.
The generator also determines the proper order of calling corresponding functions if the program fragment contains several modules or subsystems.

\begin{figure}
    \centering
    \includegraphics[scale=0.6]{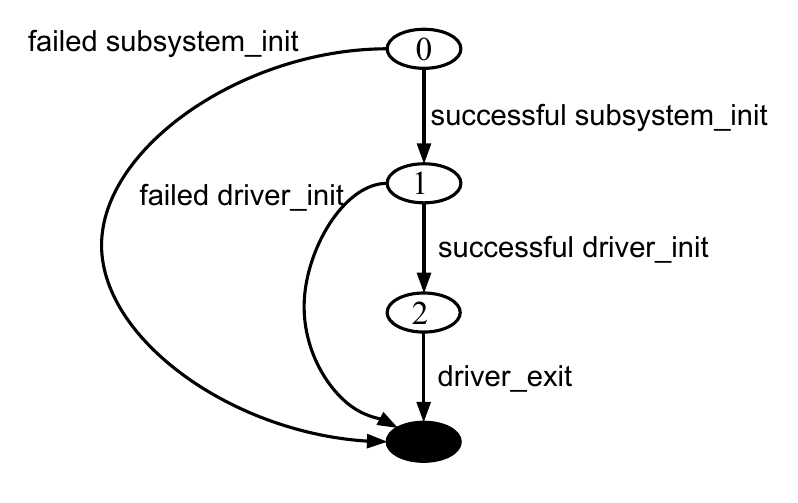}
    \caption{Order of calling subsystem initialization, driver initialization and driver exit functions.}
    \label{fig:emgsubsystems}
\end{figure}

There is an example of an order for calling a subsystem initialization and the driver's initialization and exit functions in Fig~\ref{fig:emgsubsystems}.
Both initialization functions can return an error, and curve arrows reflect these cases.
The Linux kernel always initializes subsystems before loadable kernel modules, so the corresponding function precedes the module's initialization function.

\textit{Callbacks caller} is the most complicated scenario generator. 
At the first stage, the generator collects all interface specifications. These specifications list declarations used at model callbacks' invocations.
Interfaces are grouped into categories corresponding to the specific functionality of the Linux kernel.
A category can contain the following types of interfaces:

\begin{itemize}
    \item Kernel functions include functions for registering, deregistering, creating, or destroying relevant data.
    \item Callbacks' declarations help to detect callbacks' groups implemented in a program fragment.
    \item Containers are structures storing pointers to callbacks or other important data, e.g., pointers to device tables representing supported devices and some of their characteristics.
    \item Resources are objects passed as parameters to multiple callbacks.
\end{itemize}

Then, the generator selects categories of those interfaces that are implemented or used in the source code of the program fragment.
The generator gets all the specifications that describe callbacks' invocation after that.
We refer to them as event specifications.
The specifications' notation extends the format of an intermediate model.
Main extensions are the following: new labels' attributes for their binding to particular interfaces and categories, flags to define labels as resources, callbacks, and containers.
There is a new kind of action named \textit{callback}.
It describes a callback's call with labels corresponding to resources.

Each category needs an event specification to generate an environment model.
A user can explicitly set the connection.
Otherwise, a heuristic algorithm is used to select a suitable event specification for an interface category.

The final step is the transformation of transition systems from event specifications to intermediate model notation in accordance with the interfaces implemented or used in the program fragment.
At this step, the generator performs several modifications and simplifications, such as replacing callback actions with code blocks containing explicit callback calls.
If there are several implementations of callbacks in a certain category, the generator inserts additional scenario models based on the event specification.
Also, it prints callback wrappers if they are implemented as static functions.

\textit{User model composer} accepts serialized and manually modified intermediate models.
They can be obtained at any launch of Klever from its working directory and then manually corrected.
A user can also make the intermediate model from scratch.
The generator does not make any modifications to the provided specifications.
It either adds or replaces parts of the intermediate model already received as input.


In total, the collection describes callbacks from the following categories: USB, USB Serial, IIO, PCI, Platfrom, Power Management, Class, I2C, SCSI, HID, TTY, Parport, file and seq operations, block driver, Super, Net, interrupts, kthreads, timers, tasklets and workqueues.

\begin{figure}
\centering
\begin{tikzpicture}
\begin{axis}[
    xlabel={The number of most widespread callback types},
    ylabel={The share of corresponding callback implementations},
    xmin=0, xmax=700,
    ymin=0, ymax=100,
    xtick={0,100,200,300,400,500,600,700},
    ytick={0,20,40,60,80,100},
    yticklabel={\pgfmathprintnumber\tick\%}
]
\addplot coordinates {(1,15) (2,27) (3,40) (4,45) (5,48) (10,60) (20,71) (50,82) (100,88) (200,94) (300, 96) (500, 99) (700, 100)};
\end{axis}
\end{tikzpicture}
\caption{Dependence of the share of implementations of callbacks of the most common types on the number of all callbacks implementations for Linux kernel 3.14.}
\label{figure:handlers}
\end{figure}
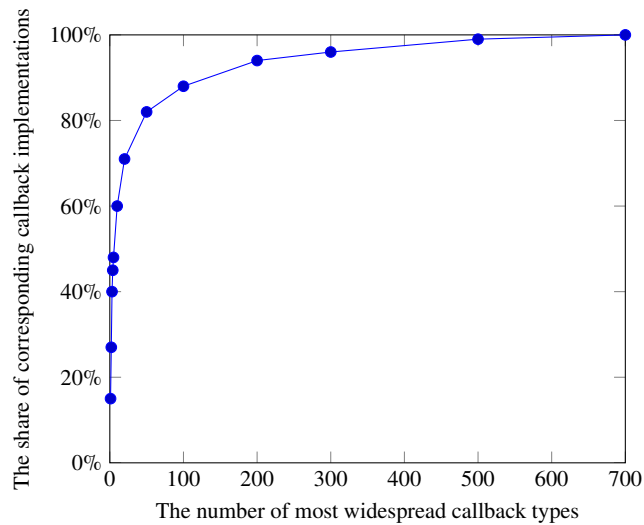

The level of code coverage and the quality of verification results depend on the number of callbacks invoked by environment models.
Fig.~\ref{figure:handlers} shows the dependence of the share of implementations of device driver callbacks of the most popular types (categories in our terminology) relative to the number of implementations of all types of callbacks for Linux 3.14.
The plot demonstrates that more than \SI{80}{\percent} of all callbacks belong to the 100 most popular categories.
Thus, it would require significantly more effort to develop environmental model specifications than we have already spent to get a significantly higher level of code coverage.
The code coverage for particular Linux kernel versions is discussed in Section~\ref{section:results}.

The Klever project's source tree contains environment model specifications in the directory \textit{presets/jobs/specifications/linux}.
It includes 36 specifications with 20 KLOC of DSL code and 1.5 KLOC of models in C.
The collection of specifications is suitable for the following Linux kernel versions: 2.6.33, 3.14, 4.6.7, 4.15, 4.17 and 5.5.
One can use these specifications for other versions as well, but the quality of verification results can decrease.


The \textit{presets/jobs/specifications} directory contains auxiliary functions for different adaptations in the \textit{verifier} and \textit{common} subdirectories.
They contain models for allocating memory, returning undefined values and so on. 

\subsubsection{Requirement Specifications}

Klever includes 34 requirement specifications for checking memory safety and rules of correct usage of the Linux kernel API as well as for finding data races in Linux kernel modules and subsystems.
In general, these requirement specifications allow to detect the following faults:
\begin{itemize}
\item Buffer overflows.
\item Null pointer dereferences.
\item Uninitialized memory usages.
\item Double or incorrect memory releases.
\item Data races and deadlocks.
\item Leaks of specific resources of the Linux kernel.
\item Incorrect function calls depending on the execution context.
\item Incorrect initialization of specific Linux kernel resources.
\end{itemize}

Requirement specifications for the Linux kernel are located in the directory \textit{presets/jobs/specifications/linux}.
The requirement specifications base is placed at \textit{presets/jobs/specifications/linux/Linux.json}.
In total, requirement specifications' size is about 5.4~KLOC.
The requirement specifications base consists of approximately 900~lines.
Development of a new typical requirements specification can take about 1~week for experts and a month for novices.

\subsection{Adaptation for BusyBox}

The BusyBox project incorporates several command-line utilities such as \textit{cat}, \textit{tar}, or \textit{tail} into a single software system.
Each utility is called an \textit{applet}.
Each applet can consist of several C source files.
During its operation, it relies only on the \textit{libbb} shared library of the project and the C standard library.
There is strictly one entry point function in an applet.
Its name is a concatenation of the applet's name and the suffix \textit{\_main}.
The function prototype coincides with the prototype of the \textit{main} function.

The adaptation in Klever matches the BusyBox 1.28.3 version. It includes one decomposition tactic and a set of requirement and environment model specifications.

\subsubsection{Decomposition Tactic}

The decomposition tactic for the BusyBox project separates the \textit{libbb} library and applets. Its algorithm includes the following steps:
\begin{enumerate}
    \item Search all C source files from the project source code.
    Distinguish files from the \textit{libbb} directory as a separate program fragment.
    \item Find those files that implement entry point functions according to the mentioned above pattern.
    We refer to them as main files.
    \item Collect C source files from other than \textit{libbb} directory that have dependencies with the main file using a callgraph.
    \item Prepare a list of program fragments for applets by joining main files with their found dependencies.
\end{enumerate}

The algorithm has an additional configuration parameter to add C source files from the \textit{libbb} directory to each program fragment.
It is enabled by default.
The adaptation has a decomposition specification for BusyBox.
It implies removing eight \textit{libbb} files from each program fragment.
The excluded files cause problems for Klever and verification tools.

Klever successfully generates program fragments for 185 applets of Busybox configured in \textit{defconfig}.

\subsubsection{Environment Modeling}

An applet's environment model calls the main function and provides stubs for certain undefined functions.
Different adaptations share a scenario generator to call functions in random order and with undefined parameters.
It is used in the BusyBox adaptation to generate calls of applets' main functions.
There is an example of such a harness for the SSL client applet in Fig.~\ref{listing:sslmain}.

\begin{figure}
\centering
\begin{minted}[frame=single,fontsize=\scriptsize]{C}
void main(void) {
    int ldv_arg_0;
    char **ldv_arg_1;
    ldv_arg_1 = external_allocated_data();
    return ssl_client_main(ldv_arg_0, ldv_arg_1);
}
\end{minted}
\caption{Example of an environment model for the SSL client applet.}
\label{listing:sslmain}
\end{figure}

The set of environment model specifications includes several types of models in the C programming language:
\begin{enumerate}
    \item Models of functions from the C standard library and POSIX such as functions for working with strings, \textit{exit}, \textit{fork}, etc.
    These models can be reused at verification of other user-space programs.
    \item Models of functions from the excluded \textit{libbb} source files.
\end{enumerate}

The general set of functions is tiny; there are no more than a dozen functions of each mentioned type.
With the current environment models, the line coverage constitutes \SI{93}{\percent} and function coverage is \SI{86}{\percent}.

\subsubsection{Requirement Specifications}



We verified memory safety and consistency of opens and closes of file descriptors and pipes for BusyBox applets.
No faults have been found.
The number of false alarms was less than a dozen.

\section{Practical Evaluation}
\label{section:results}

Klever has been used for verification of various operating system kernels and drivers.
As a proof of concept that can be publicly demonstrated, Klever helped to reveal more than 400 faults acknowledged by the Linux kernel developers\footnote{\url{http://tinyurl.com/487ntw4d}}.
Many of these bugs were reported by students of several universities and participants of the Google Summer of Code program.

We verified a subset of loadable kernel modules of Linux~5.5 released on January~26, 2020 and Linux~5.12\nobreakdash-rc3 released on March~14, 2021 to evaluate different aspects of the user experience with Klever.
We built the Linux kernel for the architecture \textit{x86\_64} and configuration \textit{allmodconfig}.
Target loadable kernel modules were considered program fragments.
On average, the size of the chosen modules was the closest to the mean size of all modules of the Linux kernel.
These modules were from the following subdirectories of directory \textit{drivers}: \textit{hid}, \textit{hwmon}, \textit{media}, \textit{mtd}, \textit{platform}, \textit{staging}, \textit{usb} and \textit{video}.
Table~\ref{tab:kernelStats} presents characteristics of these modules.
There are about \SI{4}{\percent} more modules for Linux~5.12\nobreakdash-rc3 in comparison with Linux~5.5.

\begin{table}
\caption{Characteristics of target modules.}
\label{tab:kernelStats}
\centering
\begin{tabular}{ | l  | c | c | }
\hline
\textbf{Characteristic} &
\textbf{Linux~5.5} &
\textbf{Linux~5.12\nobreakdash-rc3} \\
\hline
Number of modules &
\SI{2059}{} &
\SI{2141}{} \\
\hline
Number of C source files &
\SI{3383}{} &
\SI{3645}{} \\
\hline
Total size of modules &
2.7 MLOC &
2.8 MLOC \\
\hline
Average size of modules &
1.3 KLOC &
1.3 KLOC \\
\hline
\end{tabular}
\end{table}

We checked the given modules against 17 requirement specifications intended for checking the correct usage of the Linux kernel API\footnote{Identifiers of checked requirement specifications are as follows: \textit{alloc:\{irq, spinlock, usb lock\}}, \textit{arch:\{asm:dma-mapping, mm:ioremap\}}, \textit{drivers:\{base:class, clk1, clk2\}}, \textit{drivers:usb:\{core:driver, core:urb, core:usb:coherent, core:usb:dev, gadget:udc-core\}}, \textit{fs:sysfs:group}, \textit{kernel:\{module, rcu:update:lock\}}, \textit{net:core:dev}.}.
Also, we checked memory safety for them.
Underneath, we especially distinguish memory safety, since the solution of the corresponding verification tasks considerably differs from the solution of other verification tasks.
Klever uses a CPAchecker configuration on the base of symbolic memory graphs~\cite{Dudka:2013:BPF} for memory safety and a CPAchecker configuration on the base of block-abstraction memoization and a predicate analysis~\cite{Friedberger:2016:CPA} for requirement specifications intended for checking the correct usage of the Linux kernel API.

For experiments, we used Klever~3.1 and the verification tool CPAchecker~r36955 from branch \textit{klever-fixes}.
Klever~3.1 was released with some improvements in specifications intended specially for Linux~5.5, while for Linux~5.12\nobreakdash-rc3 we did not make any specific adjustments in both specifications and tools.

All experiments were conducted on an OpenStack virtual machine with 8~virtual cores of the Intel~Xeon~E312xx (Sandy~Bridge) CPU, 64~GB of memory and  Debian~9 (Stretch) on board.
The computational resource limits per verification task were 5~minutes of CPU time and 5~GB of memory.

\subsection{Verification Results}

Klever successfully generated \SI{31444}{} verification tasks for Linux~5.5 and \SI{31886}{}
verification tasks for Linux~5.12\nobreakdash-rc3.
Table~\ref{tab:verificationResultsStat} presents the overall statistics of obtained verification results.
One can see from the table that the \textit{Unsafes} share is approximately 50 times more for memory safety.
Besides, the relative number of \textit{Unknowns} is also 2.5 times higher.
For requirement specifications intended for checking the correct usage of the Linux kernel API the most common verdict was \textit{Safe}, i.e. the verification tool could prove correctness under the certain assumptions.

\begin{table}
\caption{Overall statistics of verification results. \textit{Unsafes} correspond to warnings issued by the verification tool (they may be either faults or false alarms). \textit{Safe} means that the verification tool could prove correctness under the certain assumptions. \textit{Unknowns} correspond to failures of the verification tool or Klever components.}
\label{tab:verificationResultsStat}
\centering
\begin{tabular}{ | l  | c | c | }
\hline
\textbf{Verdict} &
\textbf{Linux~5.5} &
\textbf{Linux~5.12\nobreakdash-rc3} \\
\hline
\multicolumn{3}{ | c | }{Memory safety} \\
\hline
Unsafe &
280 (\SI{14}{\percent}) &
278 (\SI{13}{\percent}) \\
\hline
Safe &
910 (\SI{44}{\percent}) &
901 (\SI{42}{\percent}) \\
\hline
Unknown &
869 (\SI{42}{\percent}) &
962 (\SI{45}{\percent}) \\
\hline
\multicolumn{3}{ | c | }{Correct usage of the Linux kernel API} \\
\hline
Unsafe &
85 (\SI{0.24}{\percent}) &
110 (\SI{0.3}{\percent}) \\
\hline
Safe &
\SI{29033}{} (\SI{83}{\percent}) &
\SI{29319}{} (\SI{81}{\percent}) \\
\hline
Unknown &
\SI{5885}{} (\SI{17}{\percent}) &
\SI{6968}{} (\SI{19}{\percent}) \\
\hline
\end{tabular}
\end{table}

The changes in verification results for Linux~5.12\nobreakdash-rc3 relatively to Linux~5.5 are not huge.
One can see that there are fewer \textit{Safes} and more \textit{Unknowns}.
Foremost, this is associated with some new expressions in the source code that cause failures of various Klever components and CPAchecker.
For instance, CIL started to fail on 19 modules from the directory \textit{drivers/staging/greybus/}.
The number of \textit{Unsafes} did not change too much.
A detailed analysis shows that the most noticeable changes were caused by two reasons.
The first reason is that modules started to use new APIs that were not yet modeled.
The second reason is that we reported rather many fixes of faults revealed for Linux~5.5.
They were accepted before releasing Linux~5.12\nobreakdash-rc3.
Thus, Klever does not find corresponding faults anymore.
Anyway, these slight changes demonstrate that, despite there being quite considerable changes in target programs (see Table~\ref{tab:programExamples} and Table~\ref{tab:kernelStats}), Klever still could produce comparable and valuable verification results without any changes in the project-specific adaptation for the Linux kernel.

Hereinafter, in this subsection, we consider \textit{Unsafes} and \textit{Unknowns} in detail.

\subsubsection{Unsafes}

Table~\ref{tab:bugsAndFalseAlarms} shows the number of faults and the number of false alarms.
The number of faults revealed during checking memory safety is approximately the same as the total number of faults that were found for all remaining requirement specifications.
Below, we will separately treat these faults.
A false alarm rate when checking memory safety is higher than for requirement specifications intended for checking the correct usage of the Linux kernel API.
It is explained by the fact that more accurate environment models are necessary for it.
For Linux~5.12\nobreakdash-rc3 the false alarm rate increased.
Primarily, this is due to the same reasons as for changes in the number of \textit{Unsafes}.

\begin{table}
\caption{Distribution of \textit{Unsafes}.}
\label{tab:bugsAndFalseAlarms}
\centering
\begin{tabular}{ | l  | c | c | }
\hline
\textbf{Unsafe type} &
\textbf{Linux~5.5} &
\textbf{Linux~5.12\nobreakdash-rc3} \\
\hline
\multicolumn{3}{ | c | }{Memory safety} \\
\hline
Fault &
34 (\SI{12}{\percent}) &
25 (\SI{9}{\percent}) \\
\hline
False alarm &
246 (\SI{88}{\percent}) &
253 (\SI{91}{\percent}) \\
\hline
\multicolumn{3}{ | c | }{Correct usage of the Linux kernel API} \\
\hline
Fault &
30 (\SI{35}{\percent}) &
33 (\SI{30}{\percent}) \\
\hline
False alarm &
55 (\SI{65}{\percent}) &
77 (\SI{70}{\percent}) \\
\hline
\end{tabular}
\end{table}

Table~\ref{tab:falseAlarmReasons} demonstrates the distribution of false alarms for different reasons.
Inaccurate analysis performed by CPAchecker when checking memory safety gives considerably more false alarms than for other requirement specifications.
Besides, it is more demanding for the accuracy of environment models as was already stated above.
Most such false alarms are due to the absence of definitions and models of functions that considerably influence verification results, e.g., those functions that allocate and initialize specific kernel resources.

\begin{table}
\caption{False alarm reasons.}
\label{tab:falseAlarmReasons}
\centering
\begin{tabular}{ | l  | c | c | }
\hline
\textbf{False alarm reason} &
\textbf{Linux~5.5} &
\textbf{Linux~5.12\nobreakdash-rc3} \\
\hline
\multicolumn{3}{ | c | }{Memory safety} \\
\hline
Inaccurate environment models &
123 (\SI{50}{\percent}) &
128 (\SI{51}{\percent}) \\
\hline
Inaccurate requirement specifications &
0 (\SI{0}{\percent}) &
0 (\SI{0}{\percent}) \\
\hline
Imprecise verification tool &
114 (\SI{46}{\percent}) &
117 (\SI{46}{\percent}) \\
\hline
Others &
9 (\SI{4}{\percent}) &
8 (\SI{3}{\percent}) \\
\hline
\multicolumn{3}{ | c | }{Correct usage of the Linux kernel API} \\
\hline
Inaccurate environment models &
20 (\SI{36}{\percent}) &
42 (\SI{55}{\percent}) \\
\hline
Inaccurate requirement specifications &
22 (\SI{40}{\percent}) &
25 (\SI{32}{\percent}) \\
\hline
Imprecise verification tool &
11 (\SI{20}{\percent}) &
9 (\SI{12}{\percent}) \\
\hline
Others &
2 (\SI{4}{\percent}) &
1 (\SI{1}{\percent}) \\
\hline
\end{tabular}
\end{table}

There is a noticeable difference in the distribution of false alarms for Linux~5.12\nobreakdash-rc3 in contrast to Linux~5.5 for requirement specifications intended for checking the correct usage of the Linux kernel API.
It can be explained by the fact that modules started to use new APIs that should be modeled.

Users can wonder whether it is possible to get rid of false alarms since they hinder the assessment of verification results.
Hopefully, there are some large groups of false alarms.
For example, about \SI{20}{\percent} of all false alarms related to a verifier are due to CPAchecker not supporting attributes \textit{packed} and \textit{aligned}.
Also, there are several non-modeled functions, each causing about \SI{10}{\percent} of all false alarms related to environment models.

Fixing verification tools is a very laborious job, but they get better every year.
Additionally, we improve environment models steadily.
Often, it is not enough to eliminate the initial reasons for false alarms.
After that, models for other functions may be required, or CPAchecker can produce false alarms due to imprecise analysis.

Caution: We manually assessed the reported \textit{Unsafes}, so there may be some mistakes.
If the reader tries to repeat this work, one can obtain a more or less different distribution.

\subsubsection{Faults}

In this subsection, we discuss the faults that we have found, as they represent quite an important outcome for most users.
There are two primary questions regarding the discovered faults:
\begin{itemize}
\item Whether these faults are crucial enough or benign, of which static analysis tools are often accused of.
\item Whether these faults can be easily found by other approaches.
\end{itemize}


Klever helped to identify faults that can result in various bad consequences, such as NULL pointer dereference, buffer overflow, usage of non-initialized data, memory leaks, and so on.
We prepared patches fixing some of them and sent these patches to the Linux kernel developers to demonstrate that these faults are really essential.
They accepted most of the patches and rejected just a few of them since they corresponded to situations that are not possible in practice (we need to improve our environment models).

Appendix~\ref{Appendix:Faults} enumerates those faults that were found by Klever in  Linux~5.5.
The total number of faults that were fixed by developers after our reports is 20.
The number of faults that were somehow identified by other means as well is 11.
Developers backported 14 commits to stable branches of the Linux kernel repository.
These branches represent a long-term support for older versions.
Backporting to them indicates that the Linux kernel developers agree that the corresponding faults are really significant.

Here is a discussion of the ability of other approaches to identify faults found by Klever.
Many faults (41 of 64) revealed in Linux~5.5 happen on error handling paths that are executed extremely seldom. 
Moreover, they are unlikely covered by testing and dynamic analysis tools.
Most faults (52 of 64) manifest when a complicated interaction with an environment takes place.
This assumes a specific order of callback invocation depending on their return values as well as the specific initialization of their arguments.
Besides, it may be necessary to comprehend the semantics of the called functions, e.g., when, after a successful invocation of one function, another one should be called.
These cases may be pretty hard for lightweight static analysis tools.
Though, certain tools will be able to find some faults revealed thanks to Klever after specific tuning.

We have not reported all detected faults to the Linux kernel developers yet, since this process occupies quite a lot of time.
Nevertheless, we are going to report the remaining faults gradually.

\subsubsection{Unknowns}

Table~\ref{tab:Unknowns} shows the number of \textit{Unknowns} due to failures of CPAchecker and other reasons.
The share of \textit{Unknowns} reported by CPAchecker when checking memory safety is substantially higher than the one for remaining requirement specifications.
There have been no considerable changes in the distribution of \textit{Unknowns} due to CPAchecker and other reasons over time.
Though, the absolute number of \textit{Unknowns} increased, which was already explained at the beginning of this subsection.

\begin{table}
\caption{Distribution of \textit{Unknowns}.}
\label{tab:Unknowns}
\centering
\begin{tabular}{ | l  | c | c | }
\hline
\textbf{Component} &
\textbf{Linux~5.5} &
\textbf{Linux~5.12\nobreakdash-rc3} \\
\hline
\multicolumn{3}{ | c | }{Memory safety} \\
\hline
CPAchecker &
557 (\SI{64}{\percent}) &
595 (\SI{62}{\percent}) \\
\hline
Others &
312 (\SI{36}{\percent}) &
367 (\SI{38}{\percent}) \\
\hline
\multicolumn{3}{ | c | }{Correct usage of the Linux kernel API} \\
\hline
CPAchecker &
579 (\SI{10}{\percent}) &
683 (\SI{10}{\percent}) \\
\hline
Others &
\SI{5306}{} (\SI{90}{\percent}) &
\SI{6285}{} (\SI{90}{\percent}) \\
\hline
\end{tabular}
\end{table}

Table~\ref{tab:CPAUnknowns} shows that timeouts are the most common reason for CPAchecker \textit{Unknowns} especially for memory safety.
For Linux~5.12\nobreakdash-rc3 their share decreased primarily due to more parsing failures.

\begin{table}
\caption{Distribution of CPAchecker \textit{Unknowns}.}
\label{tab:CPAUnknowns}
\centering
\begin{tabular}{ | l  | c | c | }
\hline
\textbf{Unknown type} &
\textbf{Linux~5.5} &
\textbf{Linux~5.12\nobreakdash-rc3} \\
\hline
\multicolumn{3}{ | c | }{Memory safety} \\
\hline
Timeout &
510 (\SI{92}{\percent}) &
526 (\SI{88}{\percent}) \\
\hline
Others &
47 (\SI{8}{\percent}) &
69 (\SI{12}{\percent}) \\
\hline
\multicolumn{3}{ | c | }{Correct usage of the Linux kernel API} \\
\hline
Timeout &
467 (\SI{81}{\percent}) &
467 (\SI{68}{\percent}) \\
\hline
Others &
112 (\SI{19}{\percent}) &
216 (\SI{32}{\percent}) \\
\hline
\end{tabular}
\end{table}

Users may wonder why we performed experiments with such small limits on computational resources per verification task.
For instance, for verification tasks from the \mbox{SV-COMP} benchmark suite these limits are 3 times higher~\cite{Beyer:2021:SVC}.

Lack of memory is not such a common reason of \textit{Unknowns} while with a lower limit, one can solve more verification tasks in parallel or solve them on a less powerful machine.
Timeouts happen much more often.
We evaluated the influence of increasing the CPU time limit on the number of timeouts.
Using the timeout of 15~minutes resulted in 5 more \textit{Safes} and one more \textit{Unsafe} for requirements specification \textit{drivers:clk1} while originally, there were about \SI{3300}{} of \textit{Safes} and 14 \textit{Unsafes}.
The CPU time consumption of CPAchecker rose from 15 hours to 25 hours.
Further increasing the CPU time limit gave similar improvements in verification results but  degradation in CPU time.

\subsection{Code Coverage}

Table~\ref{tab:codeCoverage} presents current code coverage.
It is reasonably high for particular subdirectories, and for some modules it is \SI{100}{\percent} while for some subdirectories and modules it is very low.
This is the case because we developed environment model specifications not for all types of modules, as was discussed in the previous section.
Code coverage for Linux~5.12\nobreakdash-rc3 varies not so much.
This again demonstrates that Klever can be applied to newer versions of target programs without considerable changes (indeed there were no changes at all for the given experiment).

\begin{table}
\caption{Code coverage.}
\label{tab:codeCoverage}
\centering
\begin{tabular}{ | l  | c | c | c | c | }
\hline
\multirow{3}{*}{\textbf{Subdirectory}} &
\multicolumn{2}{c |}{\textbf{Linux~5.5}} &
\multicolumn{2}{c |}{\textbf{Linux~5.12\nobreakdash-rc3}} \\
&
Line coverage, &
Function coverage, &
Line coverage, &
Function coverage, \\
&
KLOC &
thousands &
KLOC &
thousands \\
\hline
hid &
24/37 (\SI{64}{\percent}) &
1.2/2.1 (\SI{58}{\percent}) &
25/38 (\SI{65}{\percent}) &
1.3/2.2 (\SI{59}{\percent}) \\
\hline
hwmon &
33/63 (\SI{52}{\percent}) &
1.5/3.9 (\SI{38}{\percent}) &
34/67 (\SI{51}{\percent}) &
1.5/4.1 (\SI{37}{\percent}) \\
\hline
media &
114/354 (\SI{32}{\percent}) &
5.7/19.6 (\SI{29}{\percent}) &
120/368 (\SI{33}{\percent}) &
6.0/20.0 (\SI{29}{\percent}) \\
\hline
mtd &
19/51 (\SI{37}{\percent}) &
0.86/2.8 (\SI{31}{\percent}) &
18/50 (\SI{36}{\percent}) &
0.81/2.7 (\SI{30}{\percent}) \\
\hline
platform &
13/34 (\SI{37}{\percent}) &
0.76/2.5 (\SI{30}{\percent}) &
15/39 (\SI{38}{\percent}) &
0.89/2.9 (\SI{31}{\percent}) \\
\hline
staging &
53/159 (\SI{33}{\percent}) &
2.4/8.2 (\SI{29}{\percent}) &
40/127 (\SI{32}{\percent}) &
1.8/6.3 (\SI{28}{\percent}) \\
\hline
usb &
66/163 (\SI{40}{\percent}) &
3.5/8.8 (\SI{40}{\percent}) &
63/158 (\SI{40}{\percent}) &
3.3/8.5 (\SI{39}{\percent}) \\
\hline
video &
43/95 (\SI{45}{\percent}) &
1.8/4.6 (\SI{40}{\percent}) &
48/94 (\SI{51}{\percent}) &
1.9/4.6 (\SI{41}{\percent}) \\
\hline
\textbf{Total} &
365/957 (\SI{38}{\percent}) &
17.8/52.7 (\SI{34}{\percent}) &
364/941 (\SI{39}{\percent}) &
17.5/51.7 (\SI{34}{\percent}) \\
\hline
\end{tabular}
\end{table}

\subsection{Verification Time}

The overall time spent by Klever and the CPU time consumed by Klever components and CPAchecker is demonstrated in Table~\ref{tab:verificationTime}.

\begin{table}
\caption{Verification time in hours.}
\label{tab:verificationTime}
\centering
\begin{tabular}{ | l  | c | c | }
\hline
\textbf{Type of time} &
\textbf{Linux~5.5} &
\textbf{Linux~5.12\nobreakdash-rc3} \\
\hline
\multicolumn{3}{ | c | }{Memory safety} \\
\hline
Overall time &
10 &
11 \\
\hline
CPU time of CPAchecker &
52 &
53 \\
\hline
CPU time of Klever components &
16 &
17 \\
\hline
\multicolumn{3}{ | c | }{Correct usage of the Linux kernel API} \\
\hline
Overall time &
71 &
85 \\
\hline
CPU time of CPAchecker &
223 &
232 \\
\hline
CPU time of Klever components &
158 &
190 \\
\hline
\end{tabular}
\end{table}

In total, it took about 7.4 days to get all the verification results.
This time can be considerably reduced by using more machines or a more powerful machine for both the generation and solution of verification tasks.
One can see that more time was necessary for Linux~5.12\nobreakdash-rc3.
The primary reason for that is that there are more modules and, thus, verification tasks.
The solution of verification tasks by CPAchecker when checking memory safety consumed the most CPU time.
The generation of verification tasks by Klever components needed a rather considerable portion of the CPU time to check the correct usage of the Linux kernel API.

It is worth noting that we launched Klever several times when checking requirement specifications devoted to the correct usage of the Linux kernel API.
For that reason, some actions, like the generation of environment models, were duplicated during the generation of verification tasks.
These actions are not very time-consuming, so if one checks these requirement specifications together, there may be up to \SI{10}{\percent} improvement in the overall time and the CPU time of Klever components.

\section{Related Work}
\label{section:relatedwork}

According to our knowledge, no existing tool automates the preparation of an arbitrary C program before verification, runs verification tools, processes verification results and provides means for their further analysis and improvement.
Below, we consider several frameworks intended for the verification of specific software, such as device drivers or embedded software.

SDV is the best-known example of an application of the automatic software verification technique in practice~\cite{Ball:2011:DSM}.
It aims at checking the correct usage of the kernel API in Windows device drivers using SLAM, YOGI, and Q~\cite{Ball:2010:SSD,Lal:2014:PSD}.
There are also LDV~Tools~\cite{Zakharov:2015:CTS}, DDVerify\cite{Witkowski:2007:MCC}, and Avinux~\cite{Post:2007:ISA} intended for verification of Linux device drivers.
These frameworks are based on CBMC~\cite{Clarke:2004:TCA} and CPAchecker~\cite{Beyer:2011:CTC}.
As a result, hundreds of unrevealed faults have been found and acknowledged by developers.

CBMC is also known for various other applications, such as verification of TinyOS~\cite{Bucur:2010:SVT} and embedded software~\cite{Schlich:2009:MCC}.
Authors deliver successful case studies as a proof of concept.
There is an IDE for development of embedded software called \textit{mbeddr} that allows to automatically run CBMC to check programs under development against a predefined set of safety properties~\cite{Carlan:2016:URC}.
The IDE also provides developers with nicely arranged results.
However, \textit{mbeddr} is not intended for automated verification of programs that were not developed using it.
Another example of a successful integration of CBMC into the development process is the recent work devoted to verification of the AWS C Common library~\cite{Chong:2021:CLM}.

DC2 aims at verification of industrial software~\cite{Ivancic:2015:SSS}.
To limit the verification scope, it generates contracts relevant for finding such faults as memory leaks and array-bound overflows.
If necessary, users can improve these contracts manually.
Then DC2 runs the Varvel model checker.
However, it is an in-house NEC research project, so it is not possible to estimate its applicability to various C programs in more detail.

\section{Conclusion}
\label{section:concl}

The paper presents the ongoing work dedicated to applying automatic software verification tools for critical industrial C programs like operating system kernels and embedded software.
We described the Klever software verification framework and demonstrated the results of its practical evaluation.
We based Klever on solutions accepted by the \mbox{SV-COMP} community.
Moreover, we keep in touch with verification tool developers to cooperate and to solve the most vital problems together, discussing the interface, providing feedback and contributing generated verification tasks to the competition benchmark suite.

Our experience clearly demonstrates that automatic software verification is a very promising area since it enables finding faults in programs, many of which could be hardly detected by other software quality assurance techniques.
The more efforts that are invested in related research, development and, especially, various applications, the more awesome achievements will be reached.
We hope that one day automatic software verification techniques will become one of the best practices, along with testing and static analysis.
Moreover, we expect that for some industries, like avionics, railways, and autonomous vehicles, the use of this technique will be required by appropriate standards.

\section{Acknowledgments}

We would like to express special thanks to Prof. Dr. Alexander K. Petrenko who was a scientific supervisor of most related works, researches and theses.
Also, much gratitude is due to Alexey Khoroshilov and Vadim Mutilin for their great ideas and suggestions starting from the beginning of this project, to Pavel Shved, Alexander Strakh, Mikhail Mandrykin, Pavel Andrianov, Anton Vasilyev, Vitaliy Mordan and Denis Efremov for participating in the development of a Klever prototype as well as for improving auxiliary tools and CPAchecker, to Vladimir Gratinskiy for investing enormous efforts in the design and development of the most close to the user and large Klever component, and to numerous students and users who made various contributions.

\bibliographystyle{spmpsci}
\bibliography{main}

\appendix
\clearpage
\setcounter{section}{0}
\renewcommand\thesection{\Alph{section}}

\section{Appendix: Auxiliary Tools}\label{Appendix:AuxiliaryTools}

\subsection{Clade}\label{Appendix:Clade}
For their operation, several Klever components require a lot of information on a program under verification and its build process, for instance:
\begin{itemize}
    \item Original program source files.
    In some cases their preprocessed versions may be necessary as well.
    \item Information on build commands such as compilation, linking, assembler, archive, move, etc.
    These build commands need to be parsed, and information about their input and output files, command-line options and environment variables should be collected.
    \item List of all header dependencies for each C source file.
    \item Graphs demonstrating connections between build commands both by parent-child relations and by relations of their input and output files.
    \item Definitions and expansions of macros and macro functions.
    Definitions and declarations of functions, types and global variables.
    Source code locations for these entities and their usages.
\item A function callgraph.
\end{itemize}

Surprisingly, for different platforms as well as for different build environments, many unexpected issues need to be solved in order to support the process of collection of such information.
Reliable solutions for these issues are proprietary; thus, we decided to develop our own tool called Clade\footnote{https://github.com/17451k/clade}.

Clade can intercept all build commands that are executed during a typical build process.
Depending on a platform, it supports several different methods to do this.
On Linux and macOS, Clade intercepts and handles calls to \textit{exec} functions issued for each build command.
This is achieved by using \textit{LD\_PRELOAD} and \textit{DYLD\_INSERT\_LIBRARIES} mechanisms of the dynamic linker correspondingly.
On Windows, Clade uses a simple debugger that:
\begin{itemize}
\item Waits for process start events that correspond to the execution of build commands.
\item Pauses the build process and reads the memory of newly created processes to find and log their command-line arguments
\item Resumes the build process.
\end{itemize}

Clade includes many extensions that can parse intercepted build commands, collect information about source code, and perform various actions using the collected data.
As a result, Clade outputs a so-called \textit{build base}, which is a stand-alone folder that can be archived, easily transferred to a different computer and used there.
Clade has a command-line interface for integration with build processes and execution of its extensions as well as a stable Python API.

\subsection{CIF}\label{Appendix:CIF}

Both Clade and some Klever components need a tool for querying and weaving source code.
Since there was not an appropriate tool at that time when we needed it, we developed CIF\footnote{https://forge.ispras.ru/projects/cif}~\cite{Novikov:2013:AIA}.

CIF implements an aspect-oriented extension for the C programming language to weave so-called cross-cutting concerns into the source code\footnote{https://cif.readthedocs.io/en/latest/aoc.html}.
We use these capabilities during environment modeling and development of requirement specifications, e.g. to replace function calls with calls to models.
CIF supports source-to-source weaving that is necessary for application of verification tools.
Besides, CIF allows basic queries to source code~\cite{Novikov:2012:UAO} which are, nevertheless, powerful enough to get all information required by Clade.

CIF is based upon GCC~7.5 and supports a vast majority of GNU C extensions that can influence verification results.
At the moment, CIF does not support some specific expressions, which may result in failures during weaving and querying of the source code.

\subsection{CIL}\label{Appendix:CIL}

CIL is a source-to-source transformation tool allowing merging preprocessed C source files~\cite{Necula:2002:CIL}.
Besides, CIL performs many optimizations like removing unused functions that can simplify verification tasks rather considerably.

Klever uses CIL from Frama-C~\cite{Kirchner:2015:FSA} with a number of our own improvements and fixes developed within the VerKer project\footnote{https://forge.ispras.ru/projects/astraver/repository/framac?rev=20.0}~\cite{Efremov:2017:DVU}.
As with CIF there are some specific unsupported code constructions.

\subsection{BenchExec}\label{Appendix:BenchExec}

BenchExec~\cite{Beyer:2019:RBR} is a convenient and reliable tool to solve verification tasks using a single machine.
It is used for benchmarking verification tools within \mbox{SV-COMP} starting from the beginning.
There are following important features of BenchExec for Klever use cases:
\begin{itemize}
    \item BenchExec allows to precisely measure and limit computational resources such as CPU time and memory.
    Moreover, there is the RunExec tool that helps to run other tools in addition to verification tools.
    \item BenchExec contains wrappers for all verification tools that participate in \mbox{SV-COMP}.
    These wrappers help to run them without changing the format of verification tasks and without tool specific output processing.
\end{itemize}

\section{Appendix: Decomposition Specification Examples}
\label{Appendix:Fragments}

There are examples of decomposition specifications in Fig.~\ref{listing:decomposition_spec1} and Fig.~\ref{listing:decomposition_spec2}.
The strings "3.14" and "1.28.3" distinguish different versions of each program.
The attribute "fragments" allows to define program fragments by listing files or other program fragments explicitly.
Attributes "add to all fragments" and "exclude from all fragments" allow to add and remove files, program fragments or even functions (PFG finds a file with a function definition and adds it).

\begin{figure}
\begin{minted}[frame=single,fontsize=\scriptsize]{json-object}
"3.14": {
  "fragments": {
    "drivers/usb/serial/usbserial.ko": [
      "drivers/usb/serial/usb_debug.ko",
      "drivers/usb/serial/usbserial.ko"
    ]
  }
}
\end{minted}
\caption{Decomposition specification example for the Linux kernel.}
\label{listing:decomposition_spec1}
\end{figure}

\begin{figure}
\begin{minted}[frame=single,fontsize=\scriptsize]{json-object}
"1.28.3": {
  "exclude from all fragments": [
    "libbb/getopt32.c",
    "libbb/xfuncs_printf.c"
  ],
  "add to all fragments": [
    "libbb/wfopen.c",
    "libbb/wfopen_input.c"
  ]
}
\end{minted}
\caption{Decomposition specification example for Busybox.}
\label{listing:decomposition_spec2}
\end{figure}

\section{Appendix: Intermediate Model Example}\label{Appendix:Env}

In Fig.~\ref{listing:process_model}, one can find a fragment of an intermediate model.
It describes a single scenario model $\varepsilon = <\mathcal{V}, \mathcal{A}, \mathcal{R}, \alpha_0>$.
In Fig.~\ref{listing:c_model} there is an equivalent scenario model generated as a function in the C programming language.

Labels $\mathcal{V}$ are defined in the \textit{labels} entry.
They are escaped with $\%$ characters in C statements and logical expressions below.

The \textit{actions} entry describes each action of $\mathcal{A}$ in details.
There is the \textit{fail} base block in the intermediate model.
The \textit{condition} entry describes its precondition $\varphi$, \textit{statements} correspond to $\beta$.
\textit{reg} is a sending signal action.
Its label $l$ equals to the name of the action.
Precondition $\varphi$ is specified in the \textit{condition} entry.
Vector $\pi$ is stored in \textit{parameters} but it is empty in the example.
\textit{Comment} entries help to generate the environment model source code with hints extracted at processing and visualization of violation witnesses.

\begin{figure}
\begin{minted}[frame=single,fontsize=\scriptsize]{json}
{
    "tty/callbacks": {
        "labels": {
            "driver": {"declaration": "struct tty_driver *driver"},
            "retval": {"declaration": "int retval"},
            "ops": {"declaration": "struct tty_operations *ops",
                    "value": "__VERIFIER_nondet_pointer()"}
        },
        "actions": {
          "callbacks": {
            "comment": "Begin functions calls.",
            "parameters": []
          },
          "alloc": {
            "comment": "Allocate device.",
            "statements": ["%driver% = tty_alloc_driver(...);"]
          },
          "reg": {
            "comment": "Register callbacks.",
            "statements": ["tty_set_operations(%driver%, %ops%);",
                           "%retval% = tty_register_driver(%driver%);"],
            "condition": ["%driver% != 0"]
          },
          "unreg": {
            "comment": "Unregister callbacks.",
            "statements": ["tty_unregister_driver(%driver%);",
                           "put_tty_driver(%driver%);"],
            "condition": ["%retval% == 0"]
          },
          "fail": {
            "comment": "Failed to register callbacks.",
            "statements": ["put_tty_driver(%driver%);"],
            "condition": ["%retval% != 0"]
          },
          "skip": {
            "comment": "Failed to allocate memory for the driver.",
            "condition": ["%driver% == 0"]
          }
        },
        "process": "(!callbacks).<alloc>.(<reg>.(<unreg> | <fail>) | <skip>)"
    }
}
\end{minted}
\caption{Scenario model example.}
\label{listing:process_model}
\end{figure}

\begin{figure}
\centering
\begin{minted}[frame=single,fontsize=\scriptsize]{C}
void entry_point(void) {
    int retval;
    struct tty_driver *driver;
    struct tty_operations *ops;
    ops = __VERIFIER_nondet_pointer();

    driver = tty_alloc_driver(...);
    if (__VERIFIER_nondet_int() && driver != 0) {
        tty_set_operations(driver, ops);
        retval = tty_register_driver(driver);
        if (retval == 0) {
            tty_unregister_driver(driver);
            put_tty_driver(driver);
        } else {
            put_tty_driver(driver);
        }
    }
}
\end{minted}
\caption{Scenario model example in the C programming language.}
\label{listing:c_model}
\end{figure}

\section{Appendix: Requirement Specifications Base}\label{Appendix:ReqSpecBase}

A requirement specifications base contains high-level descriptions of requirement specifications for a given project.
These descriptions enumerate requirement specification files, select how to generate environment models, point out specific options for a chosen verification tool, etc.
Fig.~\ref{listing:reqSpecDescs} demonstrates how we do this for requirement specifications \textit{kernel:locking:rwlock}, \textit{kernel:locking:spinlock}, \textit{kernel:module}, and \textit{memory safety} for the Linux kernel.
In this example one can see templates \textit{loadable kernel modules and kernel subsystems}, and \textit{memory safety for loadable kernel modules and kernel subsystems}.
They are used to avoid duplicate options for particular requirement specifications.
Each template describes a sequence of plugins that will be invoked by Klever~Core to generate verification tasks for the verification tool step by step.
For instance, template \textit{loadable kernel modules and kernel subsystems} requires Environment Model Generator (EMG), Requirement Specification Generator (RSG), and Final Verification Task Preparator (FVTP) to be invoked.
One can specify necessary options for each plugin, e.g. common models for RSG that will be used for all requirement specifications referring this template.

Requirement specifications themselves are described in the tree form where leaves represent particular requirement specifications.
One can specify template at any level of the tree, and it will be used by all children recursively unless it will be overwritten.
For instance, requirement specification \textit{kernel:locking:rwlock} refers template \textit{loadable kernel modules and kernel subsystems}.
Besides, it says that certain model \textit{linux/kernel/locking/rwlock.c} should be used by RSG in addition to common models.
This distinguishes the given requirement specification from the other ones.

Verifier profiles referred in the figure are discussed in Appendix~\ref{Appendix:VP}.

\begin{figure}
\centering
\begin{minted}[frame=single,fontsize=\scriptsize]{json}
{
  "templates": {
    "loadable kernel modules and kernel subsystems": {
      "plugins": [
        {
          "name": "EMG",
          "options": {
            "generators options": [...],
            "translation options": {...}
          }
        },
        {
          "name": "RSG",
          "options": {
            "model compiler input file": "scripts/mod/empty.c",
            "common models": ["linux/arch/asm/atomic.c", ...]
          }
        },
        {
          "name": "FVTP",
          "options": {
            "verifier profile": "reachability",
            "verifier": {"name": "CPAchecker", "version": "trunk:31140"}
          }
        }
      ]
    },
    "memory safety for loadable kernel modules and kernel subsystems": {...}
  },
  "requirement specifications": {
    "description": "Linux requirement specifications",
    "template": "loadable kernel modules and kernel subsystems",
    "children": [
      {
        "identifier": "kernel",
        "children": [
          {
            "identifier": "locking",
            "children": [
              {
                "identifier": "rwlock",
                "plugins": [{
                  "name": "RSG",
                  "options": {"models": ["linux/kernel/locking/rwlock.c"]}
                }]
              },
              {"identifier": "spinlock", ...}
            ]
          },
          {
            "identifier": "module",
            "plugins": [{
              "name": "RSG",
              "options": {"models": ["linux/kernel/module.c"]}
            }]
          }
        ]
      },
      {
        "identifier":  "memory safety",
        "template": "memory safety for loadable kernel modules and kernel subsystems"
      },
      ...
    ]
  }
}
\end{minted}
\caption{Requirement specifications base example.}
\label{listing:reqSpecDescs}
\end{figure}

\section{Appendix: Verifier Profile Example}\label{Appendix:VP}

There is a verifier profiles example in Fig.~\ref{listing:verifierProfiles}.
The snippet contains:
\begin{itemize}
    \item Three templates \textit{CPAchecker common}, \textit{CPAchecker BAM}, and \textit{CPAchecker SMG} that describe common CPAchecker configurations. BAM stands for Block Abstraction Memoization and SMG for Symbolic Memory Graphs.
    \item Two verifier profiles \textit{reachability}, and \textit{memory safety checking} that refer these templates and describe final configurations for CPAchecker~\textit{trunk:31440}.
\end{itemize}
It is allowed to inherit templates and to modify or add some additional options both in templates and verifier profiles.
For instance, template \textit{CPAchecker BAM} inherits template \textit{CPAchecker common}, and specifies certain safety properties to be checked as well as some extra options in addition.
Also, one is capable to provide different sets of options for various versions of the same tool by describing appropriate verifier profiles.

\begin{figure}
\centering
\begin{minted}[frame=single,fontsize=\scriptsize]{json}
{
  "templates": {
    "CPAchecker common": {
      "description": "Common options for the CPAchecker tool",
      "add options": [
        {"-setprop": "counterexample.export.exportExtendedWitness=true"},
        ...
      ]
    },
    "CPAchecker BAM": {
      "description": "CPAchecker with BAM for reachability checking",
      "inherit": "CPAchecker common",
      "safety properties": [
        "CHECK( init({entry_point}()), LTL(G ! call(__VERIFIER_error())) )"
      ],
      "add options": [
        {"-ldv-bam": ""},
        ...
      ]
    },
    "CPAchecker SMG": {
      "description": "CPAchecker SMG for memory safety checking",
      "inherit": "CPAchecker common",
      "safety properties": [
        "CHECK( init({entry_point}()), LTL(G valid-free) )",
        ...
      ],
      "add options": [
        {"-smg-ldv": ""},
        ...
      ]
    }
  },
  "profiles": {
    "reachability": {
      "CPAchecker": {
        "trunk:31140": {"inherit": "CPAchecker BAM"}
      }
    },
    "memory safety checking": {
      "CPAchecker": {
        "trunk:31140": {"inherit": "CPAchecker SMG"}
      }
    }
  }
}
\end{minted}
\caption{Verifier profiles example.}
\label{listing:verifierProfiles}
\end{figure}

\section{Appendix: Faults Found by Klever in the Linux Kernel}\label{Appendix:Faults}

Table~\ref{tab:fixedFaults} enumerates those faults that were found by Klever in Linux~5.5 and that were already fixed in the mainstream as of May 29, 2021.

\begin{table}
\caption{Faults found by Klever in Linux~5.5}
\label{tab:fixedFaults}
\centering
\begin{tabular}{ | l  | l | c | c | c | c | }
\hline
\textbf{Loadable kernel module} &
\textbf{Requirements} &
\textbf{Fixed by} &
\textbf{Reported} &
\textbf{Mainline} &
\textbf{Backported} \\
&
\textbf{specification} &
\textbf{developers} &
\textbf{by us} &
\textbf{commit\tablefootnote{Commit SHAs in Git repository https://git.kernel.org/pub/scm/linux/kernel/git/stable/linux.git.}} &
\\
\hline
aspeed-pwm-tacho.ko &
memory safety &
No &
Yes &
bc4071aafcf4 &
Yes \\
\hline
bttv.ko &
memory safety &
Yes &
No &
7b817585b730 &
Yes \\
\hline
vpif\_capture.ko &
memory safety &
No &
Yes &
602649eadaa0 &
Yes \\
\hline
qcom-camss.ko &
memory safety &
No &
Yes &
f45882cfb152 &
Yes \\
\hline
s5p-jpeg.ko &
memory safety &
No &
Yes &
0862d95b437b &
No \\
\hline
vicodec.ko &
memory safety &
Yes &
No &
f36592e7b343 &
Yes \\
\hline
rc-core.ko &
memory safety &
No &
Yes &
3b4cfc6966ec &
No \\
\hline
usbtv.ko &
memory safety &
Yes &
No &
bf65f8aabdb3 &
Yes \\
\hline
plat-ram.ko &
memory safety &
No &
Yes &
8c293f545419 &
No \\
\hline
goldfish\_pipe.ko &
memory safety &
Yes &
No &
43c2cc2864bc &
No \\
\hline
kpc2000.ko &
memory safety &
Yes &
No &
b17884ccf29e &
Yes \\
\hline
kpc\_dma.ko &
memory safety &
No &
Yes &
8ce8668bfb64 &
No \\
\hline
rts5208.ko &
memory safety &
No &
Yes &
11507bf9a883 &
Yes \\
\hline
net2272.ko &
memory safety &
Yes &
Yes &
9b719c7119e7 &
No \\
\hline
net2280.ko &
memory safety &
Yes &
Yes &
f770fbec4165 &
No \\
\hline
usbtest.ko &
memory safety &
Yes &
No &
28ebeb8db770 &
Yes \\
\hline
da8xx-fb.ko &
memory safety &
No &
Yes &
80a00e90dede &
No \\
\hline
savagefb.ko &
memory safety &
No &
Yes &
e8d35898a78e &
Yes \\
\hline
sm712fb.ko &
memory safety &
No &
Yes &
19e55a87ad62 &
No \\
\hline
rtl2832\_sdr.ko &
kernel:module &
No &
Yes &
ce5d72b6f5a0 &
No \\
\hline
isif.ko &
arch:mm:ioremap &
No &
Yes &
6651dba2bd83 &
No \\
\hline
vpss.ko &
arch:mm:ioremap &
No &
Yes &
9c487b0b0ea7 &
Yes \\
\hline
goku\_udc.ko &
arch:mm:ioremap &
No &
Yes &
0d66e04875c5 &
Yes \\
\hline
net2272.ko &
arch:mm:ioremap &
No &
Yes &
ae90cc8237bf &
No \\
\hline
sm712fb.ko &
arch:mm:ioremap &
No &
Yes &
bcee1609ba96 &
No \\
\hline
usbtv.ko &
drivers:usb:core:usb:dev &
Yes &
No &
bf65f8aabdb3 &
Yes \\
\hline
stm32\_fmc2\_nand.ko &
drivers:clk1 &
Yes &
No &
71d1f1d5958f &
No \\
\hline
vf610\_nfc.ko &
drivers:clk1 &
No &
Yes &
cb7dc3178a98 &
Yes \\
\hline
cadence-quadspi.ko &
drivers:clk1 &
Yes &
No &
c61088d1f993 &
No \\
\hline
ad5933.ko &
drivers:clk2 &
Yes &
No &
da7de29bb171 &
No \\
\hline
ci\_hdrc\_usb2.ko &
drivers:clk2 &
Yes &
No &
c2de37b31f17 &
No \\
\hline
\end{tabular}
\end{table}

\end{document}